\documentclass[twocolumn,tighten,times]{aastex631}
\usepackage{amsmath}

\newcommand{\um}{$\mu$m}
\DeclareMathOperator{\atantwo}{atan2}

\shorttitle{The magnetic fields of starburst galaxies. I}
\shortauthors{Lopez-Rodriguez, E.}
\graphicspath{{./}{figures/}}

\begin{document}

\title{The magnetic fields of starburst galaxies. I. \\
Identification and characterization of the thermal polarization in the galactic disk and outflow.}

\correspondingauthor{Enrique Lopez-Rodriguez}
\email{elopezrodriguez@stanford.edu}

\author[0000-0001-5357-6538]{Enrique Lopez-Rodriguez}
\affiliation{Kavli Institute for Particle Astrophysics \& Cosmology (KIPAC), Stanford University, Stanford, CA 94305, USA}

\begin{abstract}
Far-infrared polarized emission by means of magnetically aligned dust grains is an excellent tracer of the magnetic fields (B-fields) in the cold phase of the galactic outflows in starburst galaxies. We present a comprehensive study of the B-fields in three nearby ($3.5$--$17.2$\,Mpc) starbursts (M82, NGC~253, and NGC\,2146) at $5$\,pc--$1.5$\,kpc resolutions using publicly available $53$--$890$\,\um~imaging polarimetric observations with SOFIA/HAWC+, JCMT/POL-2, and ALMA. We find that the polarized spectral energy distributions (SEDs) of the full galaxies are dominated by the polarized SEDs of the outflows with dust temperatures of $T_{\rm{d,outflow}}^{\rm{PI}}\sim45$\,K and emissive index of $\beta_{\rm{outflow}}^{\rm{PI}}\sim2.3$. The disks are characterized by low $T_{\rm{d,disk}}^{\rm{PI}}=[24,31]$\,K and $\beta_{\rm{disk}}^{\rm{PI}}\sim1$. We show that disk- and outflow-dominated galaxies can be better distinguished by using polarized SEDs instead of total SEDs. We compute the $53$--$850$\,\um~polarization spectrum of the disk and outflow and find that dust models of the diffuse ISM can reproduce the fairly constant polarization spectrum of the disk, $\langle\,P_{\rm{disk}}\rangle=1.2\pm0.5$\%. The dust models of heterogenous clouds and two temperature components are required to explain the polarization spectrum of the outflow ($2$--$4$\% at $53$\,\um, $\sim1$\% at $850$\,\um, and a minimum within $89$--$154$\,\um). We conclude that the polarized dust grains in the outflow arise from a dust population with higher dust temperature and emissivities than those from the total flux. The B-fields of the outflows have maximum extensions within $89$--$214$\,\um~reaching heights of $\sim4$\,kpc, and flatter polarized fluxes than total fluxes. The extension of the B-field permeating the circumgalactic medium increases with increasing the star formation rate.
\end{abstract}




\section{Introduction} \label{sec:INT}

Starburst galaxies expel gas, metals, and dust from their disk to the circumgalactic medium (CGM) via their galactic outflows \citep[e.g.,][]{Veilleux2005,Veilleux2020}. If the galactic outflows contain ordered B-fields, one would expect to uncover in them the signature of polarized emission from magnetically aligned dust grains. These dust grains are aligned by radiative alignment torques \citep[RATs; e.g.,][]{HL2016}. As light propagates through a gas-filled medium with these aligned elongated dust grains, preferential extinction of radiation along one plane leads to a measurable polarization in the transmitted and emitted radiation, a process called dichroic absorption and emission. The short axes of dust grains align with the local B-field, which can be measured at optical and near-infrared (NIR) wavelengths. For polarized thermal emission observed at far-infrared (FIR) and sub-mm wavelengths, the observed position angle ($PA$) of polarization traces the orientation of the local B-field after a $90^{\circ}$ rotation

Starburst galaxies were observed using optical polarimetric observations first by \citet{Elvius1962a} and \citet{Elvius1962b} and followed up by \citet{Bingham1976,Neininger1990,Scarrott1991,Fendt1998,Yoshida2019}. The optical polarimetric observations of M\,82 showed large polarization fractions, $P>16$\%, with an azimuthal $PA$ of polarization up to $\sim2$\,kpc and centered at the core. Low $P$, $<2$\%, with a $PA$ of polarization parallel to the galaxy's disk were observed at distances $>1.5$\,kpc from the core. These polarization measurements are the signatures of a) dust scattering in the central $\sim2$\,kpc, which does not provide any information about B-fields, and b) absorptive polarization by magnetically aligned dust grains in the galaxy’s disk at distances $>1.5$\,kpc. The latter suggests the presence of a galactic mean-field dynamo parallel to the disk. However, these observations do not provide information about the B-fields in the galactic outflows.

As the scattering cross-section declines much faster, $\lambda^{-4}$, than absorption, $\lambda^{-1}$, from optical to NIR wavelengths \citep{JW2015}, NIR polarimetric observations of several starbursts were also performed to attempt the detection of B-fields along the galactic outflows \citep[e.g.,][]{Scarrott1993,Jones2000,MC2014}. Indeed $1.65$ and $2$\,\um\,polarimetric observations confirmed the contribution of a large-scale B-field in the galaxy’s disk of M\,82 \citep{Jones2000}. However, these observations were also highly contaminated by dust scattering. A naive model assuming a centrosymmetric pattern was removed from the NIR observations of M\,82. This approach revealed the signature of a B-field perpendicular to the galaxy’s disk. As the polarized extinction signature is very weak compared to the polarization induced by scattering even at these wavelengths, the residuals may still be contaminated by dust scattering. In addition, another starburst galaxy, NGC\,1808, was also observed showing a pure azimuthal pattern in the $PA$ of polarization \citep{Scarrott1993}. These results indicate that dust scattering is the dominant mechanism at optical wavelengths and that the aforementioned removal approach may not be able to be applied to all starbursts. Because dust scattering dominates at these wavelengths, B-fields along the galactic outflows in starburst galaxies are therefore very challenging to study using optical/NIR polarimetric observations.

Radio polarimetric observations have historically provided most of our knowledge about extragalactic magnetism due to widespread access to radio facilities \citep{BW2013,Beck2019}. The $3$--$20$\,cm wavelength range is sensitive to the synchrotron emission arising from the warm and diffuse phase of the ISM and is affected by Faraday rotation. Although ordered B-fields are found to be parallel to the plane of spiral normal galaxies with an X-shape away from the galaxy disk at several kpc-scales \citep{Heesen2011,Krause2020}, the origin of these B-fields is still unclear. Starburst galaxies have the strongest B-field strengths of $>\,50\,\mu$G \citep{LB2013,Adebahr2017}, assuming equipartition between B-fields and cosmic rays.  A revised version of the equipartition taking into account energy losses in starburst galaxies computed B-field strengths of $72$--$770\,\mu$G \citep{LB2013}. For M\,82, radio polarimetric observations show a dominant magnetized bar parallel to the galaxy disk, with only a hint of the B-field in the galactic outflow in the northern region \citep[labeled as 3 by][]{Adebahr2017}. This result may be caused by the short lifetime of cosmic rays (CR) due to the dense medium and the high B-field strengths of the outflow \citep{Thompson2006,Adebahr2017}. Because radio polarimetric observations suffer from Faraday rotation and a short CR lifetime, the B-fields in the galactic outflows are very challenging to characterize.

\begin{figure*}[ht!]
\includegraphics[width=\textwidth]{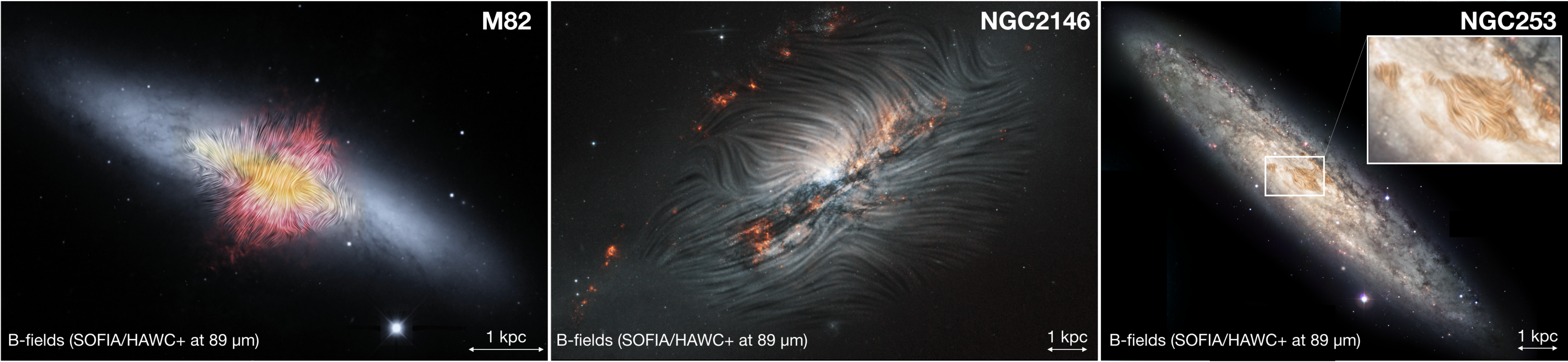}
\caption{B-field morphology of the starburst galaxies studied in this work. The B-field orientation (streamlines) using SOFIA/HAWC+ at $89$~\um~are displayed over a combination of optical and infrared images. Each panel shows the physical scale of $1$ kpc. Credits: M\,82: NASA/SOFIA/E. Lopez-Rodriguez and L. Proudfit; NASA/Spitzer/J. Moustakas et al.;  NGC\,253: ESO and NASA/SOFIA/E. Lopez-Rodriguez and L. Proudfit; NGC\,2146: ESA/Hubble \& NASA/E. Lopez-Rodriguez and L. Proudfit. \label{fig:fig0}}
\end{figure*}

Recent $50$--$850$\,\um~imaging polarimetric studies have reported polarized emission by magnetically aligned dust grains in several starburst galaxies (M\,82, NGC\,253, NGC\,2146; see Fig. \ref{fig:fig0}) \citep{Jones2019, LR2021,Pattle2021,SALSAIV}. These observations were performed using the High-resolution Airborne Wideband Camera Plus (HAWC+) onboard the 2.7-m Stratospheric Observatory for Infrared Astronomy (SOFIA) and POL-2 on the James Clerk Maxwell Telescope (JCMT). The FIR polarization arises from the thermal emission of magnetically aligned dust grains tracing a density-weighted medium along the line-of-sight (LOS) and within the beam. The Survey of extrAgaLactic magnetiSm with SOFIA (SALSA) Legacy Program measured that the $53$--$214$\,\um~polarimetric observations are sensitive to the dense ($\log_{10}(N_{\rm~HI+H_{2}}~[\rm{cm}^{-2}])=[20,23]$) and cold ($T_{\rm~d}=[20,50]$\,K) component of the ISM in galaxies \citep{SALSAIV}. These studies showed that the B-field orientation of M\,82 at $89$ and $850$ \um~is predominantly parallel to the galactic outflow with a secondary component parallel to the galaxy disk at galactocentric radius $>1$\,kpc. The B-fields in the outflows have also been reported \citep{LR2021,SALSAIV,SALSAV} within the range of $53$--$214$ \um~for M\,82 and NGC\,2146. These results are quite extraordinary because they show that magnetically aligned dust grains are present in the circumgalactic medium (CGM) around starburst galaxies up to scales of several kpc. However, a comprehensive characterization of the B-fields in the galactic outflows as a function of dust and galactic properties has still not been performed.

Our goal is to perform a systematic analysis of the B-fields in the cold phase of the galactic outflows in starburst galaxies using FIR-sub-mm polarimetric observations. This first manuscript of the series focuses on the separation of components (i.e., disk and outflow) based on an analysis of the B-field geometry and the characterization of the polarization properties of the disk and outflow. Furthermore, we compute the spectral energy distributions (SEDs) and polarization spectra of the full galaxies, outflows, and disks within the $53$--$890$ \um~wavelength range. We characterize their variations due to changes in their physical properties (i.e., dust temperature, dust grain composition, and random B-fields) and compare them with dust models of the diffuse ISM and star-forming regions of the Galaxy. Finally, we perform an energy budget of the galactic outflows and compare the extension of the B-fields in the circumgalactic region with the global star formation rate (SFR) of the starbursts. We describe the observations of the starburst galaxies in Section \ref{sec:OBS}, present the mathematical methods in Section \ref{sec:MET}, and show results in Section \ref{sec:RES}. The discussions are described in Section \ref{sec:DIS}, and our main conclusions are summarized in Section \ref{sec:CON}.

\begin{deluxetable*}{lcccccccp{4cm}}
\centering
\tablecaption{Galaxy Sample. \emph{Columns, from left to right:} 
(a) Galaxy name, 
(b) galaxy distance, 
(c) physical scale, 
(d) physical size of the beam of the observations,
(e) galaxy type, 
(f) the inclination of the galaxy, 
(g) position angle of the long axis of the galaxy in the plane of the sky, 
(h) star formation rate,
(i) references associated with the distance, inclination, tilt angles, and star formation rate. 
\label{tab:table1} 
}
\tablecolumns{6}
\tablewidth{0pt}
\tablehead{\colhead{Galaxy} & 	\colhead{Distance$^{1}$}  & \colhead{Scale} & \colhead{$\theta_{\rm{pc,beam}}$} & \colhead{Type$^{\star}$} & 
\colhead{Inclination$^{2}$ ($i$)} &	\colhead{Tilt$^{2}$ ($\theta$)} &  \colhead{SFR$^{3}$} &  \colhead{References} \\ 
 	&  \colhead{(Mpc)}	& \colhead{(pc/\arcsec)}	& \colhead{(pc)} &
\colhead{($^{\circ}$)} & \colhead{($^{\circ}$)} & \colhead{($^{\circ}$)} & \colhead{(M$_{\odot}$ yr$^{-1}$)} \\
\colhead{(a)} & \colhead{(b)} & \colhead{(c)} & \colhead{(d)} & \colhead{(e)} & \colhead{(f)} & \colhead{(g)} & \colhead{(h)} & \colhead{(i)}} 
\startdata
M\,82 		&	$3.85$	&	$18.49$	& $90$--$337$	&  I0/Sbrst		&	$76\pm1$		&	$64\pm1$		&	$13$ &
$^{1}$\citet{Vacca2015}; $^{2}$\citet{Mayya2005}; $^{3}$\citet{FS2003}		\\
NGC\,253 	&	$3.50$	&	$16.81$	 & $5$--$229$	&	SAB(s)c/Sbrst	&	$78.3\pm1.0$	&	$52\pm1$		&	$3$ &
$^{1}$\citet{RS2011}; $^{2}$\citet{Lucero2015}; $^{3}$\citet{Bolatto2013}		\\
NGC\,2146 	&	$17.20$	&	$82.61$	&  $401$--$1503$	 &	SB(s)ab/Sbrst	&	$63\pm2$		&	$140\pm2$	&	$20$ &
$^{1}$\citet{Tully1988}; $^{2}$\citet{Tarchi2004}; $^{3}$\citet{Gorski2018} 		\\
\enddata
\tablenotetext{{\star}}{Galaxy type from NASA/IPAC Extragalactic Database (NED; \url{https://ned.ipac.caltech.edu/})}
\end{deluxetable*}

\section{Galaxy sample and archival data} \label{sec:OBS}

We analyze the starburst galaxies with publicly available FIR and sub-mm polarimetric observations. Table \ref{tab:table1} shows the properties of the galaxies used in this work. 

FIR polarimetric observations were taken from the SALSA\footnote{Data can be obtained at the SALSA website \url{http://galmagfields.com/}} Legacy Program \citep{SALSAIII,SALSAIV}. M\,82 and NGC\,2146 polarimetric observations were performed with SOFIA/HAWC+ at $53$, $89$, $154$, and $214$ \um~with beam sizes (FWHM) of $4.85\arcsec$, $7.8\arcsec$, $13.6\arcsec$, and $18.2\arcsec$, and pixel scales of $2.55\arcsec$, $4.02\arcsec$, $6.90\arcsec$, $9.37\arcsec$ (i.e., Nyquist sampling). NGC~253  polarimetric observations were performed at $89$ and $154$ \um. All observations were performed using the on-the-fly-mapping observing mode with total execution times in the range of $[1.0,2.4]$h and reduced by \citet{SALSAIII}. 

Sub-mm polarimetric observations of M\,82 were taken with JCMT/POL-2 at $850$ \um~with a beam size of $14\arcsec$, pixel scale of $4.0\arcsec$, and an execution time of $12.5$h previously published by \citet{Pattle2021}. We resampled the data at a pixel scale of $7.5\arcsec$ (i.e., Nyquist sampling). Note that the original data are oversampled, so this resampling does not increase the signal-to-noise ratio (SNR) of the final data products. 

Sub-mm polarimetric observations of NGC\,253 were obtained using ALMA polarimetric mode at $890$ \um~($336.5$ GHz; Band 7) with a beam size of $0.32\arcsec \times 0.28\arcsec$ at a position angle of $-88.8^{\circ}$, pixel scale of $0.073\arcsec$, and an execution time of $2.4$h  (PI: Hughes, A., ID:2018.1.01358.S; Belfiori, D. et al. in preparation). We resampled the data at a pixel scale of $0.15\arcsec$ (i.e., Nyquist sampling). Note that the original data are oversampled, so this resampling does not increase the SNR of the final data products. 

\begin{figure*}[ht!]
\centering
\includegraphics[width=0.9\textwidth]{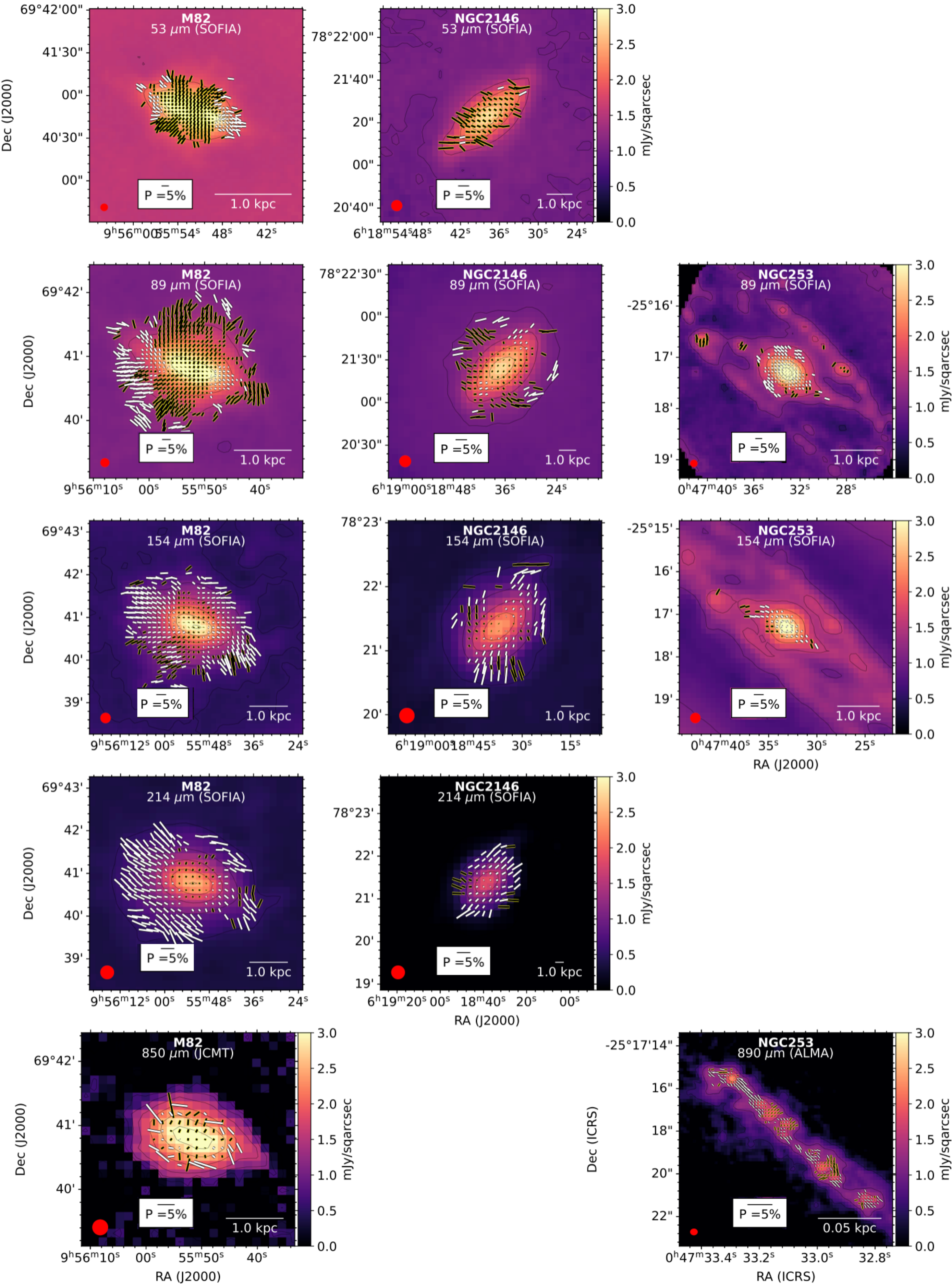}
\caption{B-field orientation and polarization maps of the FIR and sub-mm polarimetric observations of starburst galaxies. The total intensity (colormap in log-scale) maps of the starburst galaxies M\,82 (left), NGC\,2146 (middle), and NGC\,253 (right) for the FIR ($53-214$ \um; rows 1-4) and sub-mm ($850$ and $890$ \um; last row) polarimetric observations are displayed.  The B-field orientation of the disk (white lines) and outflow (yellow-black lines) with their lengths proportional to the polarization fraction are displayed. Only polarization measurements with $PI/\sigma_{\rm{PI}} \ge 2$, $P\le 20$\%, and $I/\sigma_{\rm{I}} \ge 60$ are displayed. The contour levels start at $\log_{10}{I} = 0$ and increase in steps of 0.5. Each panel shows the legends of the physical scale, a $5$\% polarization fraction, and the beam size (red circle).  \label{fig:fig1}}
\end{figure*}

\begin{figure*}[ht!]
\centering
\includegraphics[width=0.9\textwidth]{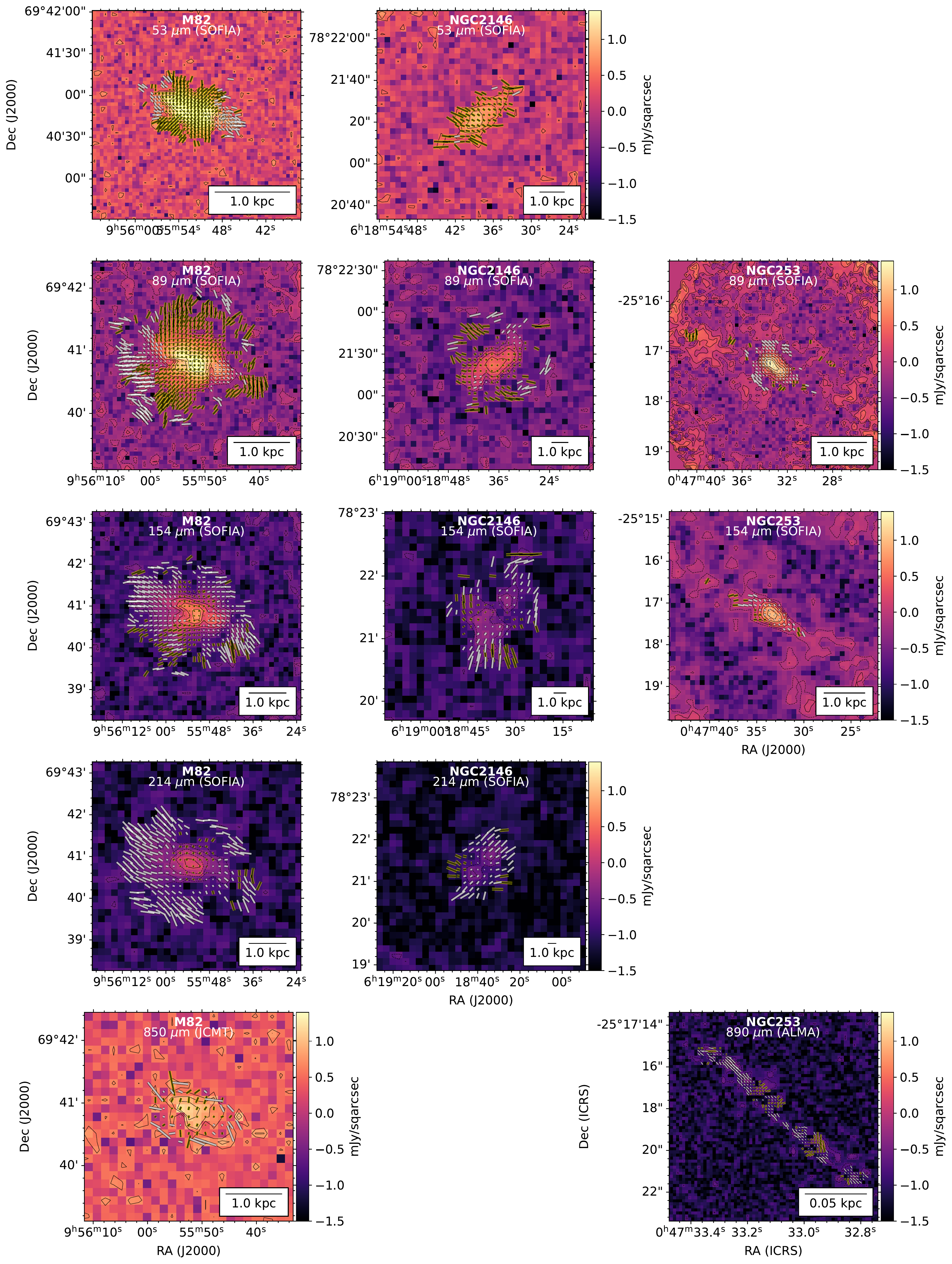}
\caption{B-field orientation and polarization maps of the FIR and sub-mm polarimetric observations of starburst galaxies. The polarized intensity (colormap in log-scale) maps of the starburst galaxies M\.82 (left), NGC\,2146 (middle), and NGC\,253 (right) for the FIR ($53-214$ \um; rows 1-4) and sub-mm ($850$ and $890$ \um; last row) polarimetric observations (top of each panel) are displayed.  The B-field orientation of the disk (white lines) and outflow (yellow lines) with their lengths proportional to the polarization fraction are displayed. Only polarization measurements with $PI/\sigma_{\rm{PI}} \ge 2$, $P\le 20$\%, and $I/\sigma_{\rm{I}} \ge 60$ are displayed. The contour levels start at $\log_{10}{PI} = 3\sigma_{\rm{PI}}$ and increase in steps of 0.2. A legend showing the physical scale of the image is shown in each panel.
 \label{fig:fig2}}
\end{figure*}

We show the B-field orientation over the total (Figure \ref{fig:fig1}) and polarized (Figure \ref{fig:fig2}) intensity maps of the starburst galaxies used in this work. The polarization measurements were rotated by $90^{\circ}$ to show the B-field orientation and the lengths are proportional to the polarization fraction. Only polarization measurements with $PI/\sigma_{\rm{PI}} \ge 2$, $P\le 20$\%, and $I/\sigma_{\rm{I}} \ge 60$ are displayed, where $\sigma_{\rm{PI}}$ and $\sigma_{\rm{I}}$ are the uncertainties of the polarized, $PI$, and total, $I$, intensities, respectively.

\section{Methods}\label{sec:MET}

We describe the methodology to estimate the resolved and unresolved polarization measurements and the approach to disentangle the B-field orientation of the disk and outflow regions. This section also describes the fitting procedure for the total and polarized SEDs.

\subsection{Polarization measurements}\label{subsec:Pmes}

To account for the vector quantity of the polarization measurements, we estimate the polarization fraction and B-field orientation of the resolved observations as follows. We estimate the mean of the polarization fraction of the individual polarization measurements, $\langle P^{\rm{hist}} \rangle$, per galaxy and per band as

\begin{equation}\label{eq:Phist}
\langle P^{\rm{hist}} \rangle = \frac{\sqrt{\langle Q^2 + U^2 - \sigma_{Q}\sigma_{U}\rangle}}{\langle I \rangle}
\end{equation}
\noindent
where $\langle \rangle$ represents the mean of the selected polarization measurements, and $\sigma_{Q}$ and $\sigma_{U}$ are the uncertainties of the Stokes $Q$ and $U$, respectively. The $1\sigma$ uncertainty of $\langle P^{\rm{hist}} \rangle$ is estimated as the standard deviation of the distribution of the individual measurements.

We also compute the integrated polarization fraction because it may be scientifically useful for observations of unresolved starburst galaxies. We estimate the integrated polarization fraction, $P^{\rm{int}}$, as

\begin{equation}\label{eq:Pint}
P^{\rm{int}} = \frac{\sqrt{\langle Q \rangle^2 + \langle U \rangle^2 - \langle \sigma_{Q} \rangle \langle \sigma_{U} \rangle}}{\langle I \rangle}
\end{equation}
\noindent
where $\langle \rangle$ is the mean of the Stokes $IQU$, and $\langle \sigma_{Q} \rangle$ and $\langle \sigma_{U} \rangle$ are the standard deviation of the mean of the Stokes $Q$ and $U$, respectively.

The difference between $\langle P^{\rm{hist}} \rangle$ and $P^{\rm{int}}$ is that the latter is the average polarization fraction of an unresolved galaxy weighted with larger polarized intensity, $PI$. The uncertainty is a measurement of the accuracy of the polarization fraction of an unresolved galaxy. $\langle P^{\rm{hist}} \rangle$ is the estimation of the average polarization fraction of all the analyzed independent beams across the galaxy, not weighted by $PI$. The uncertainty is a measurement of the dispersion of polarization fractions within the analyzed distribution of individual measurements.

\subsection{Disentangle the disk and outflow components}\label{subsec:MET_PA}

\begin{figure*}[ht!]
\includegraphics[width=\textwidth]{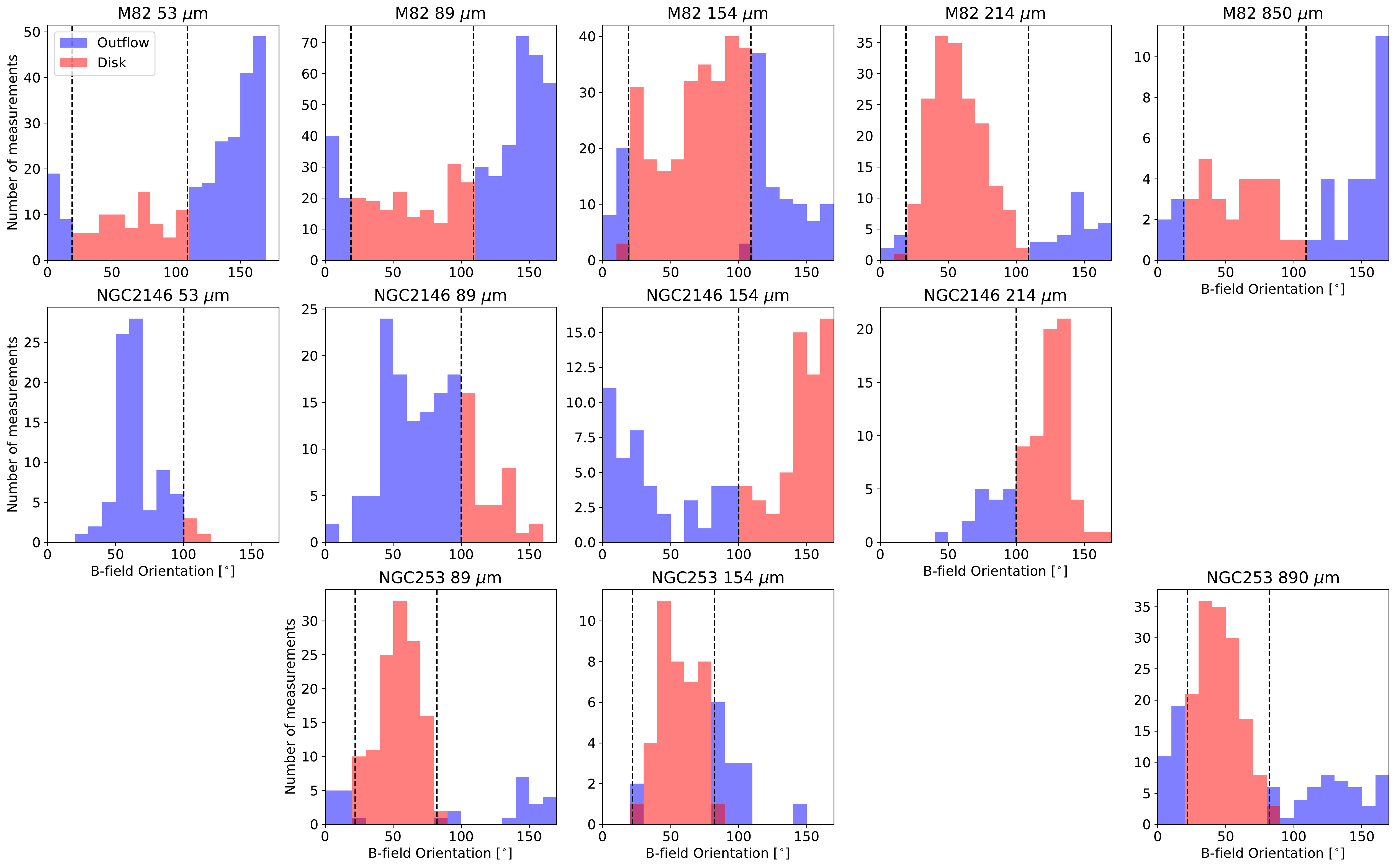}
\caption{Histograms of the B-field orientation of the outflow and disk of starburst galaxies. The histograms of the B-field orientation of the outflow (blue) and disk (red) in $10^{\circ}$ bins, as well as, the $PA_{\rm{B},-}$ and $PA_{\rm{B},+}$ from Eq. \ref{eq:PA_max_min} are shown. The spatial location of the polarization measurements associated with the outflow and disk are shown in Figure \ref{fig:fig1}. 
 \label{fig:fig3}}
\end{figure*}

We disentangle the B-field orientations from the disk and outflow of each starburst galaxy using a geometric analysis. We perform the following steps:

1. We select the polarization measurements using the following quality cuts: $PI/\sigma_{\rm{PI}} \ge 2$, $P\le 20$\%, and $I/\sigma_{\rm{I}} \ge 60$. 

2. We fit the distribution of measurements of the Stokes $Q$ and $U$ of each galaxy using a Gaussian profile with two free parameters: the mean and standard deviation, $\sigma$. 

3. We estimate the range of B-field orientations, $PA_{\rm{B}}$, using the mean and standard deviation of the best fit Gaussian profile, $\sigma$, values of the Stokes $Q$ and $U$ computed in step 2. Specifically, the range of maximum, $PA_{\rm{B},+}$, and minimum, $PA_{\rm{B},-}$, B-field orientation around the mean are estimated as

\begin{equation}\label{eq:PA_max_min}
PA_{\rm{B}, \pm} = \frac{1}{2}\atantwo \left(\langle Q \rangle\pm\sigma_{Q}, \langle U \rangle\pm\sigma_{U} \right)
\end{equation}
\noindent
where $\langle Q \rangle$ and $\langle U \rangle$ are the mean of the Stokes $Q$ and $U$ of the Gaussian profiles, and $\sigma_{Q}$ and $\sigma_{U}$ are the standard deviation of the Gaussian profiles, respectively. Note that $PA_{\rm{B},\pm}$ provides the B-field orientation, rather than the polarization angle.

4. We compute the angular difference between the tilt of the galaxy, $\theta$, and the mean of the B-field orientations within and outside the $PA_{\rm{B}, \pm}$ range. Specifically, $PA_{\rm{B,1}}$ is the mean within the $PA_{\rm{B}, \pm}$  range and $PA_{\rm{B,2}}$ is the mean outside the $PA_{\rm{B}, \pm}$ range. We estimate $\Delta B_{1} = | \theta - PA_{\rm{B,1}}|$ and $\Delta B_{2} = | \theta - PA_{\rm{B,2}}|$.

5. Finally, we select the larger angular difference, $\Delta B_{1}$ or $\Delta B_{2}$, as the outflow region and the smaller as the disk region.

\subsection{Total and Polarized SED Fitting}\label{subsec:MET_SEDFitting}

The $50-250$ \um\  total and polarized SEDs are characterized using a single-temperature modified blackbody function expressed as

\begin{equation}\label{eq:mBB}
F_{\nu} (M_{\rm d},\beta,T_{\rm d}) = \frac{M_{\rm d}}{D_{\rm L}^{2}}k_{\lambda_{0}}\left(\frac{\nu}{\nu_{0}}\right)^{\beta}B_{\nu}(T_{\rm d})
\end{equation}
\noindent
where M$_{\rm d}$ is the dust mass, $D_{\rm L}$ is the luminosity distance to the source, $k_{\lambda_{0}}$ is the dust mass absorption coefficient to be $0.29$ m$^{2}$ kg$^{-1}$ at a wavelength of $\lambda_{0} = 250$ \um\ \citep{Wiebe2009}, $\beta$ is the dust emissive index, and $B_{\nu}(T_{\rm d})$ is the blackbody function at a characteristic dust temperature of $T_{\rm d}$.

We have three free model parameters: $M_{\rm d}$, $\beta$, and $T_{\rm d}$ that we fit within the $50-250$ \um\ wavelength range. We compute a Markov Chain Monte Carlo (MCMC) approach using the No-U-Turn Sampler \citep[NUTS;][]{NUTS} method in the \textsc{python} code \textsc{pymc3} \citep{pymc}. The prior distributions are set to flat within the range of $\log_{10}(M_{\rm{d}} [M_{\odot}]) = [2,9]$, $\beta = [1,3]$, and $T_{\rm d} = [10,100]$ K. We run the code using 5 chains with 10,000 steps and a 2,000 burn-in per chain, which provides 50,000 steps for the full MCMC code useful for data analysis.


\begin{figure*}[ht!]
\includegraphics[width=\textwidth]{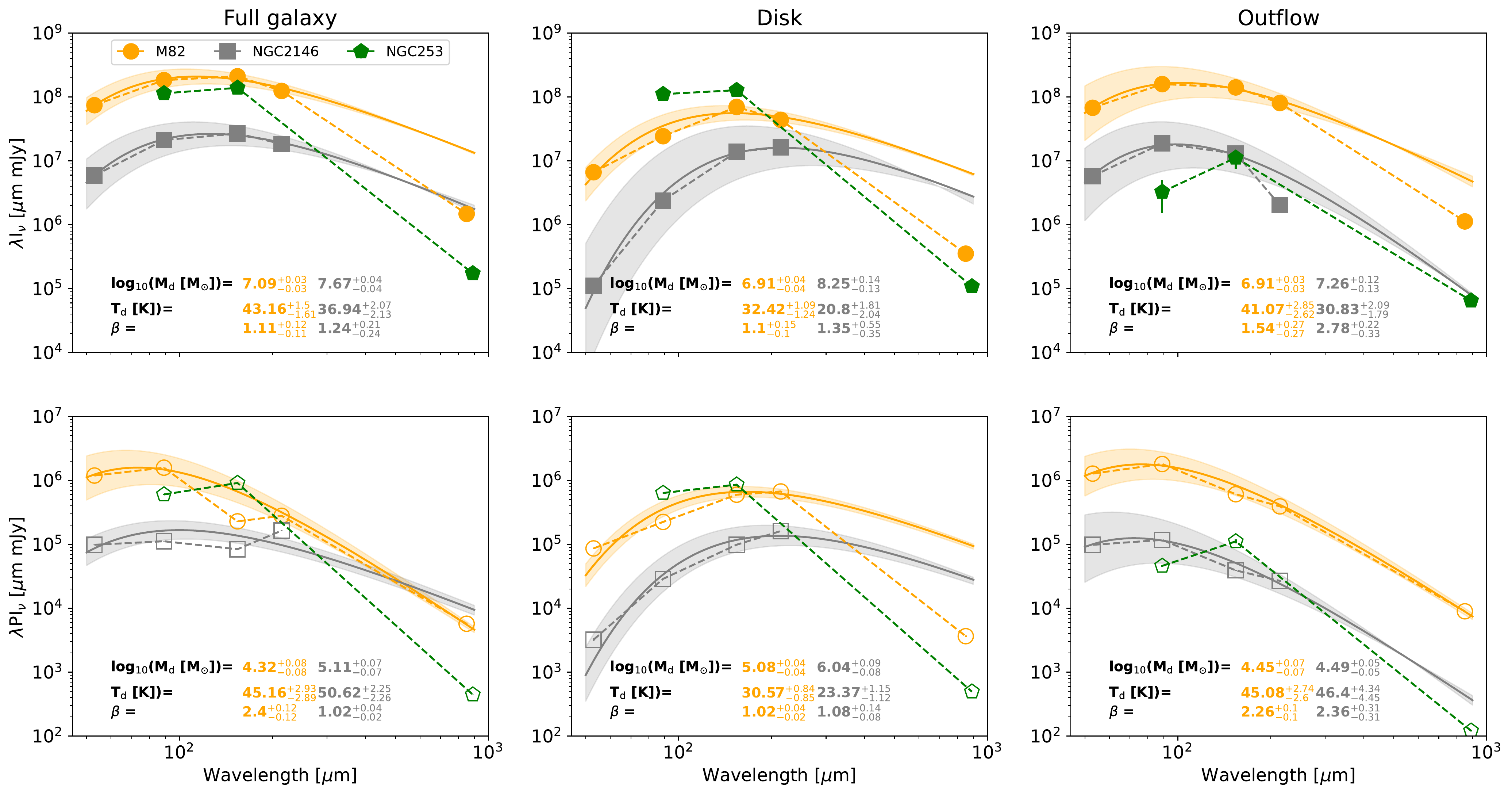}
\caption{Total and polarized SEDs of the full starburst, and only disk and outflow regions. The total ($\lambda I_{\nu}$; dashed lines and filled symbols; top panels) and polarized ($\lambda PI_{\nu}$; dashed lines and open symbols; bottom panels) SEDs for the full galaxy (left), disk (middle), and outflow (right) are shown. The best fit (solid line) and $1\sigma$ uncertainty (shadowed region) of the modified blackbody functions with their characteristic dust temperature, mass, and $\beta$ are shown in each panel.
 \label{fig:fig4}}
\end{figure*}

\section{Results}\label{sec:RES}

Figure \ref{fig:fig1} shows the B-field orientation and polarization maps of the disk and outflow regions per band of the starburst galaxies (Table \ref{tab:table1}). We disentangle the outflow and disk regions following the methodology described in Section \ref{subsec:MET_PA}. Given the complexity of the B-field orientation at some wavelengths, we use the observations at the shorter wavelength common for all starbursts, i.e., $89$ \um. At $89$ \um, the distribution of B-field orientations is dominated by the galactic outflow in M\,82 and NGC\,2146, and by the disk in NGC\,253. The best fit using a Gaussian profile with the mean and $1\sigma$ uncertainty to the measured Stokes $Q$ and $U$ (step 2 in Section \ref{subsec:MET_PA}) are shown in Appendix \ref{App:StokesQU} (Figure \ref{fig:fig1App}). We use the range of B-field orientation at $89$ \um~to separate the outflow and disk range at all wavelengths. We show the histograms of the B-field orientation of the outflow and disk in Figure \ref{fig:fig3}, which corresponds to the results of steps 3-5 in Section  \ref{subsec:MET_PA}. The $PA_{\rm{B},\pm}$ range is shown as vertical dashed lines. In the following sections, we analyze the total and polarized SEDs and polarized spectra of the full galaxy, disk, and outflow regions. This section ends with an analysis of the polarization measurements across the galactic outflows.

\begin{figure*}[ht!]
\includegraphics[width=\textwidth]{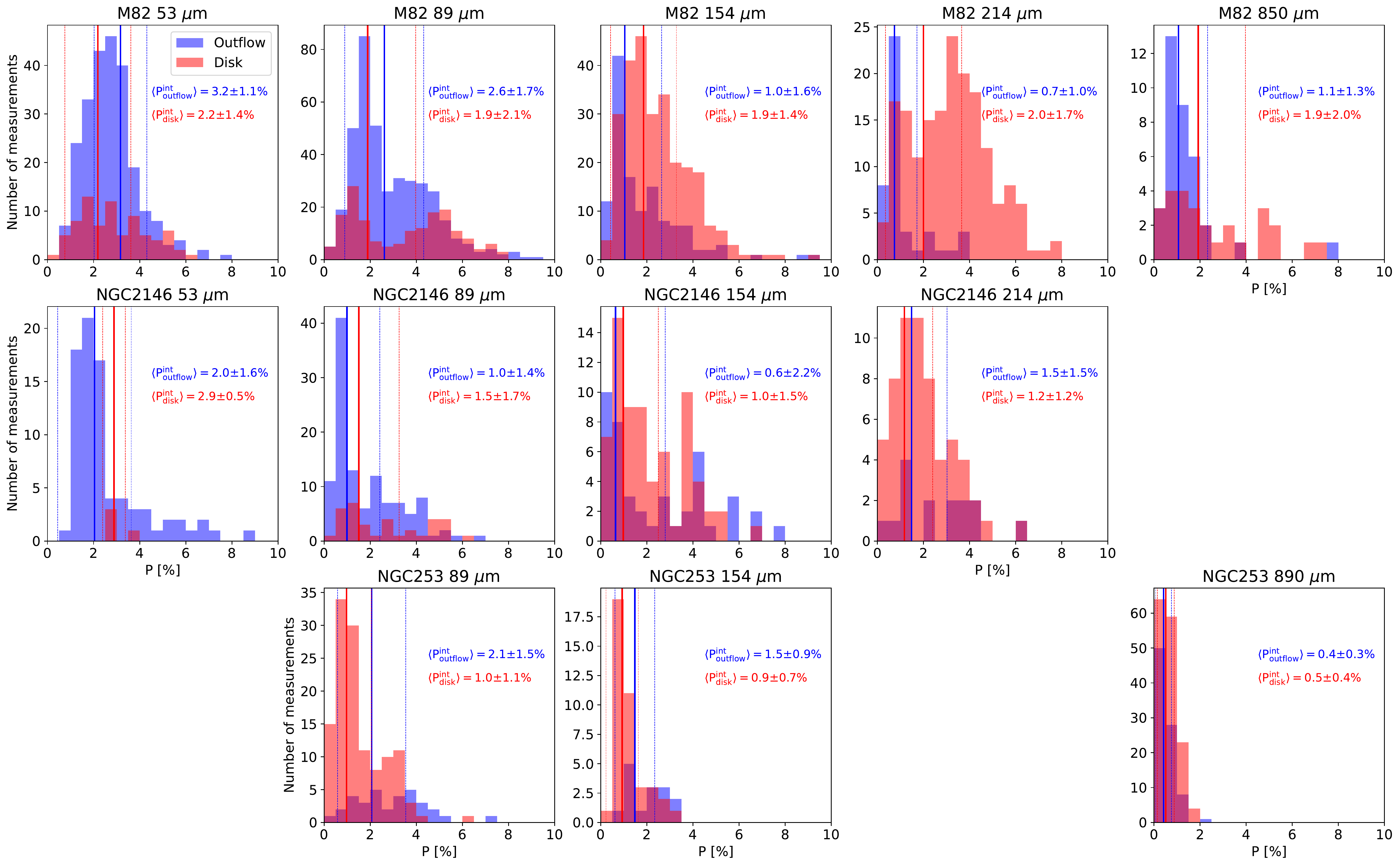}
\caption{Histograms of the polarization fraction of the outflow and disk of starburst galaxies. We use the polarization fraction of the outflow (blue) and disk (red) from the same regions shown in Figures \ref{fig:fig1} and \ref{fig:fig3}. The histograms are computed in $0.5$\% bins. The median and $1\sigma$ dispersion of the histogram are shown in each panel. 
 \label{fig:fig5}}
\end{figure*}

\subsection{Total flux SEDs}\label{subsec:RES_I}

We compute the integrated SEDs of the full galaxy, disk, and outflow regions (Figure \ref{fig:fig4}). The total flux is estimated as the sum of the Stokes $I$ using the polarization measurements shown in Figure \ref{fig:fig1}. All the polarization measurements were used for the full galaxy. Only those labeled as disk and outflow were used for the disk and outflow, respectively. We characterize the total SEDs by fitting a modified blackbody function in the $50-250$ \um~wavelength regime as described in Section \ref{subsec:MET_SEDFitting}. The sub-mm photometric measurements were not included in the fitting because the large-scale fluxes are missing in both JCMT and ALMA observations. In addition, the ALMA observations have a substantially better angular resolution, $0.3\arcsec$, and cover a smaller region, $\sim150$ pc, than those observations at FIR wavelengths, $4-18\arcsec$, with maps of several kpc. If the sub-mm observations are included the modified blackbody function tends to obtain $\beta>3$. We do not perform the fitting procedure in NGC\,253 because this galaxy only has two measurements in the $50-250$ \um~wavelength range. Figure \ref{fig:fig4} shows the best fit of the modified blackbody functions and the associated $1\sigma$ uncertainty for each source and region.

We find that the galactic outflow and disk have different total flux SEDs. The galactic outflow is characterized by having larger dust temperatures, T$_{\rm{d,outflow}}^{\rm{I}} = [31,41]$ K, and $\beta$, $\beta_{\rm{outflow}}^{\rm{I}} = [1.54,2.80]$, than those in the disk, T$_{\rm{d,disk}}^{\rm{I}} = [21,32]$ K and $\beta_{\rm{disk}}^{\rm{I}} = [1.10,1.35]$. The larger temperature in the outflow is expected due to the heating from the starburst activity. The larger $\beta$ may be a combination of a) the presence of temperature gradients in the outflows, and b) selection bias due to the decrease of polarization measurements toward larger wavelengths (Section \ref{subsec:RES_r}). Note that the $850$ \um~total flux of M\,82 is well below the expected total flux from the best-fit modified black body function using the $50-250$ \um~wavelength range. This is due to a potential loss of large-scale flux by the observations at $850$ \um~JCMT/POL-2 and that the extended host galaxy is barely detected at $850$\,\um~in comparison to the $50-250$ \um~SOFIA/HAWC+ observations (Section \ref{subsec:DIS_OBS}).

The total flux SEDs of the full galaxy have dust temperatures slightly larger than the individual outflow and disk regions, and $\beta$ tends to have values close to $1$. This result is due to the combined contribution of both outflow and disk components. Together they flatten the SED and make it appear hotter than the individual components, with a $\beta$ similar to that from the disk. Thus, the outflow is not well-characterized if the total flux SED of the full galaxy is used. This is not the case when using the polarized SEDs (Section \ref{subsec:RES_PI}).

\subsection{Polarized flux SEDs}\label{subsec:RES_PI}

The polarized SEDs are fitted using the procedure described in  Section \ref{subsec:MET_SEDFitting}, which is the same procedure used for the total flux SEDs. Only the $50-250$ \um~wavelength range is fitted. We do not perform the fitting procedure in NGC\,253 because this galaxy only has two measurements in the $50-250$ \um~wavelength range. Figure \ref{fig:fig4} shows the best fit of the modified blackbody functions and the associated $1\sigma$ uncertainty for each source and region.

We find that the disk and outflow polarized SEDs have different shapes. The disk is characterized by having low dust temperatures, $T_{\rm{d,disk}}^{\rm{PI}} = [24,31]$ K, and emissivities, $\beta_{\rm{disk}}^{\rm{PI}} \sim 1$. The dust temperature and $\beta$ are similar, within the uncertainties, to the total flux SED in the disk. The outflow has high dust temperatures,  $T_{\rm{d,outflow}}^{\rm{PI}} \sim 45$ K, and emissivities, $\beta_{\rm{outflow}}^{\rm{PI}} \sim 2.3$. We measure that the dust temperature is higher in the polarized SED than in the total flux SED. 

Note that the best fit to the polarized SED of M\,82 also reproduces the $850$\,\um~polarized measurement of the outflow, although the $850$\,\um~was not used for the fitting procedure. This result indicates that the polarized emission arising from the $850$ \um~observations is within the beam of the observations, i.e., the polarized flux at $850$\,\um~of M\,82 from the JMCT/POL-2 observations is not lost. This is not the case when using the total flux SEDs (Section \ref{subsec:RES_I}).

We find that the modified blackbody function describing the polarized SED of the full galaxy is similar to that from the outflow region for M\,82. If the $214$\,\um~polarization measurement of NGC\,2146 is not used, the fit of the full galaxy is also similar to that from the outflow. The outflow region of NGC\,2146 at $214$\,\um~is only detected by $2-3$ independent beams, which biases the polarized SED. Figure \ref{fig:fig2} shows that the polarized flux is predominantly arising from the core of the starburst regions at all wavelengths. The disk is characterized by low surface polarized brightness arising from cold dust. We find that the polarized SED of the full galaxy can be used to characterize the starburst region even when the starburst galaxy is unresolved.

\subsection{The polarization spectra of the disk, outflow, and full starburst galaxy}\label{subsec:RES_P}

\begin{figure*}[ht!]
\includegraphics[width=\textwidth]{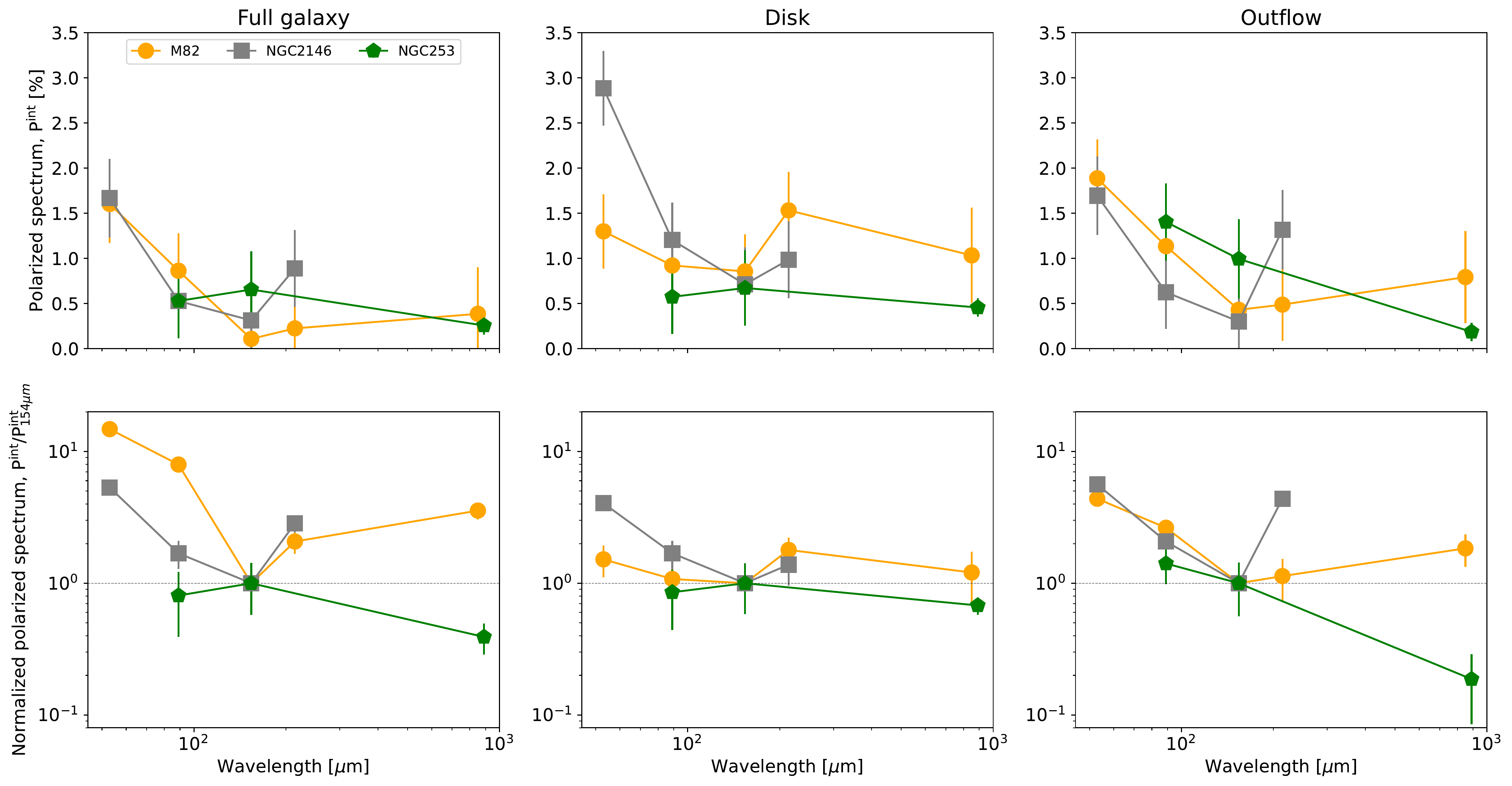}
\caption{The polarization spectra of starburst galaxies. The integrated polarization spectra ($P^{\rm{int}}$; top) for the full galaxy (left), disk (middle), and outflow (right) of M\,82 (orange), NGC\,253 (green), and NGC\,2146 (grey) are shown. The normalized polarization spectra ($P^{\rm{int}}/P^{\rm{int}}_{154\,\mu m}$; bottom) for the same regions and galaxies are shown. 
 \label{fig:fig6}}
\end{figure*}

\begin{figure*}[ht!]
\includegraphics[width=\textwidth]{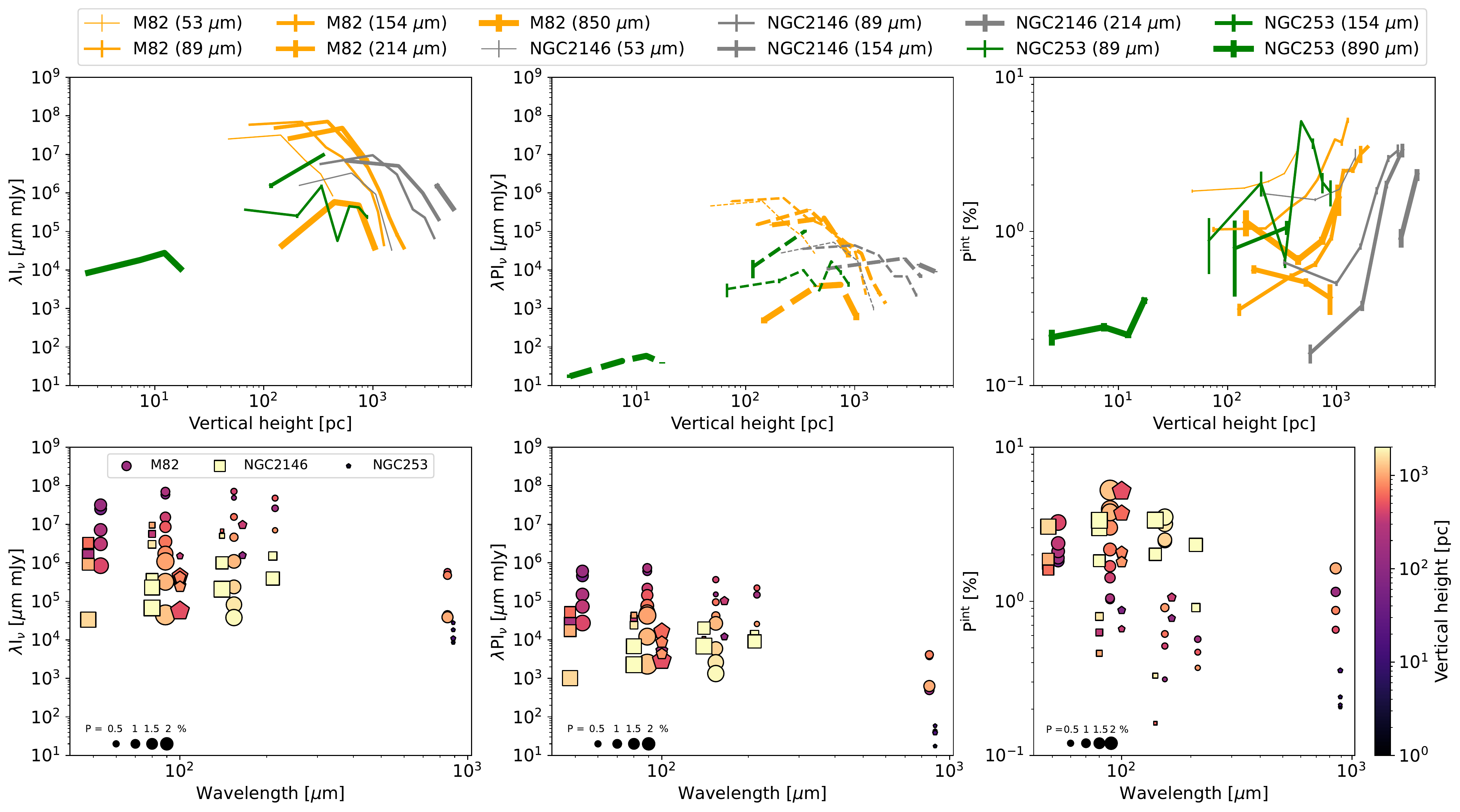}
\caption{Polarization measurements as a function of the vertical height across the outflow. 
\textit{Top}: The total flux  (left), polarized flux (middle), and polarization fraction (right) as a function of the vertical height. The linewidths increase with the wavelength for each galaxy, as shown in the top legend. 
\textit{Bottom}: The SEDs of the total flux (left), polarized flux (middle), and polarization fraction (right). The color scale changes with the vertical height, as shown in the color bar. The marker size increases with the polarization fraction, as shown in the bottom left legend.
 \label{fig:fig7}}
\end{figure*}

We compute the polarization spectra for the full galaxy, disk, and outflow regions. Figure \ref{fig:fig5} shows the histograms of the individual polarization measurements for the outflow and disk regions shown in Figures \ref{fig:fig1} and \ref{fig:fig3}. The mean polarization fraction for each region is estimated using Eq. \ref{eq:Phist} and the values can be found in Appendix \ref{App:Pmes} (Table \ref{tab:table1App}). We also estimate the integrated polarization fraction, $P^{\rm{int}}$, for the full galaxy, disk, and outflow regions using Eq. \ref{eq:Pint}, and the values can be found in Appendix \ref{App:Pmes} (Table \ref{tab:table1App}). 

Figure \ref{fig:fig6} shows the polarization spectra of the integrated polarization for the full galaxy, disk, and outflow regions. This figure also shows the normalized spectrum, $P^{\rm{int}}/P^{\rm{int}}_{\lambda_{0}}$, to the reference wavelength of $\lambda_{0}= 154$\,\um~for each region (full galaxy, disk, and outflow) of the starburst galaxies. The normalized polarization spectra minimize the effects of opacity, LOS, and POS variations of the measured B-field orientations \citep{H1999}. We find that the normalized polarization spectrum of $\langle P^{\rm{hist}} \rangle$ is similar to that of the $P^{\rm{int}}$. Hereafter, we only describe the normalized polarization spectra of $P^{\rm{int}}$. The reason is that the $1\sigma$ uncertainties of $P^{\rm{int}}$ represent the accuracy of the polarization measurement rather than the dispersion of the measurements within a region estimated for $\langle P^{\rm{hist}} \rangle$.

We find that the disk and outflow regions show different polarization spectra. For the disk region, the normalized polarization spectrum of each galaxy is fairly constant within the $50-890$\,\um~wavelength range. We estimate the median polarization fraction for the disk of starburst galaxies to be $\langle P^{\rm{int}}_{\rm{disk}}\rangle = 0.9\pm0.6$\% in the $50$--$890$\,\um~wavelength range. The larger difference is found at $53$\,\um~by NGC\,2146 because this galaxy only has three independent measurements associated with the disk. These sparse polarization measurements may be biasing the polarized spectrum toward larger polarization fractions at $53$\,\um. Although there is a significant difference in angular resolutions between the  HAWC+ ($>10\arcsec$) and the ALMA ($\sim0.3\arcsec$) polarimetric observation of NGC\,253, the thermal polarization fraction of the disk is low, $\le0.5$\%, and flat in the $50$--$890$\,\um~wavelength range.

For the outflow, the normalized polarization spectrum of each galaxy tends to decrease with increasing wavelength. The mean polarization fraction decreases from $1.8\pm0.6$\% at 53\,\um~to $0.5\pm0.3$\% at sub-mm ($>850$\,\um) wavelengths. The larger differences are found at $214$\,\um~and sub-mm ($>850$ \um) wavelengths. At $214$\,\um, NGC\,2146 has a large polarization associated with the outskirts ($\sim4$\,kpc) of the galactic outflow (Section \ref{subsec:RES_r}). At sub-mm ($>850$\,\um) wavelengths, the variation in polarization fraction is due to a) the large angular difference ($14\arcsec$ at $850$\,\um~vs.~$\sim0.3\arcsec$ at $890$\,\um), b) the distance of the polarization measurement to the midplane of the disk traced by each observation (Section \ref{subsec:RES_r}), and c) the different column density and dust temperatures traced at both FIR and sub-mm wavelengths.
 
For the full galaxy, the normalized polarization spectrum of each galaxy shows the behavior of a combined disk and outflow spectrum. At $\lambda<154$\,\um, the polarization spectrum decreases with wavelength, which is mainly dominated by the outflow. At $\lambda>154$\,\um, the polarization spectrum of all galaxies increases with increasing wavelength for the $>10\arcsec$ angular resolution observations and decreases with wavelength for the sub-arcsecond angular resolution observations.

\subsection{Outflow: vertical height dependence}\label{subsec:RES_r}

We estimate the dependence of the polarization measurements as a function of the vertical height across the outflow of starburst galaxies. We perform the following steps:

1. We construct a two-dimensional map of vertical heights from the plane of the galaxy. We define the map to be centered at the peak emission at each wavelength and rotate the map on the plane of the sky using the tilt, $\theta$, of the galaxy (Table \ref{tab:table1}). The map is rotated as $h^{\prime} = R_{z}[\theta]h$, where $h$ is the original vertical height map, $h^{\prime}$ is the new vertical height, and $R_{z}[\theta]$ is the rotation matrix for the tilt angle along the $z$-axis assuming that the galaxy's disk is on the $x-y$ plane. The map is initially in pixels, where zero corresponds to the plane of the galaxy and positively increases above and below it.  

2. We select the Stokes $IQU$ as a function of the vertical height in slots of two pixels wide. The width of two pixels corresponds to the beam size of the observations, which provides the minimum size at which each measurement is statistically independent as a function of the vertical height. 

3. The integrated polarization measurements are estimated as described in Section \ref{subsec:Pmes}. The vertical heights in pixels are converted to physical distances in pc using the scale factor in Table \ref{tab:table1}.

\begin{figure*}[ht!]
\includegraphics[width=\textwidth]{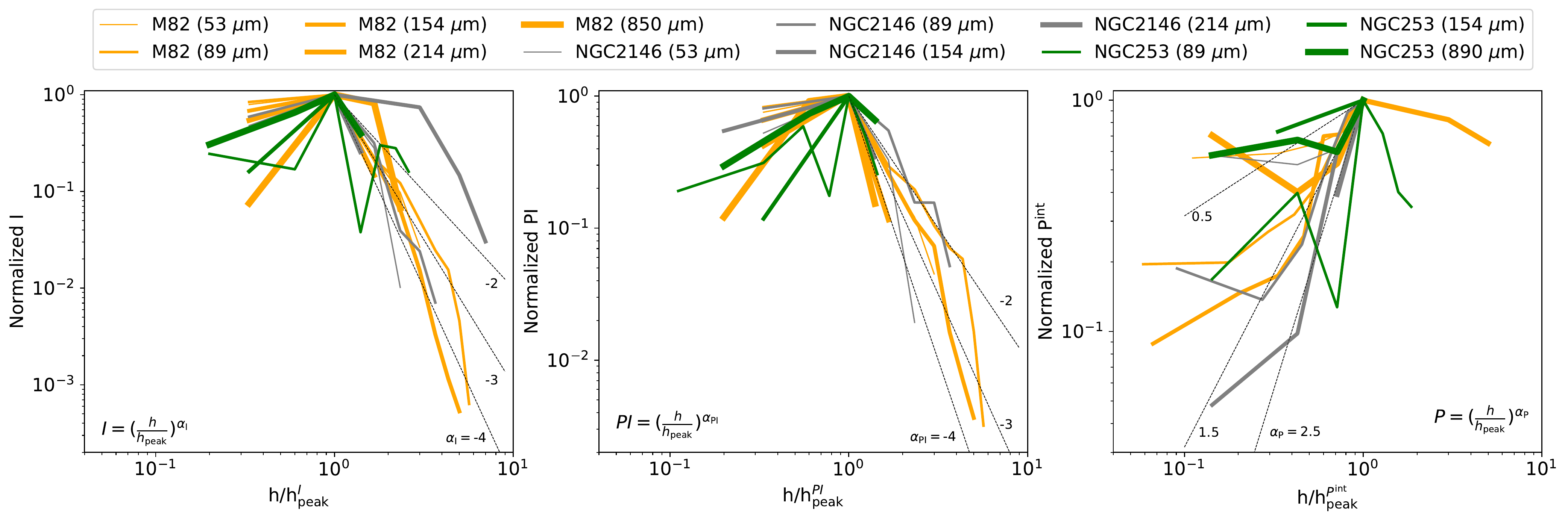}
\caption{Normalized polarization measurements of the outflow as a function of the normalized vertical height. 
The normalized total flux (left), polarized flux (middle), and integrated polarization fraction (right) are shown as a function of the normalized vertical height at the peak of the total flux, $h/h_{\rm{peak}}^{I}$, polarized flux, $h/h_{\rm{peak}}^{PI}$, and integrated polarization fraction, $h/h_{\rm{peak}}^{P^{\rm{int}}}$. The vertical heights at the peak of the polarized flux and integrated polarization fraction are shown in Appendix \ref{App:Pmes} (Table \ref{tab:table1App}). The linewidths increase with the wavelength for each galaxy, as shown in the top legend. Several power laws as indicated in each panel are shown.
 \label{fig:fig8}}
\end{figure*}

Figure \ref{fig:fig7} shows the total flux, polarized flux, and integrated polarization fraction as a function of the vertical height for the starburst galaxies. We find that the total and polarized flux decreases as the vertical height increases after the central $1$--$3$ beams per band. The increase of the total and polarized fluxes within the central two beams is because of the flux distribution across the opening angle of the outflow (Figure \ref{fig:fig2}). To study the trend of the fluxes with the vertical height, we normalize the total and polarized fluxes to the height at the peak of the total flux, $h/h_{\rm{peak}}^{I}$, and polarized flux, $h/h_{\rm{peak}}^{PI}$ for each band, respectively (Figure \ref{fig:fig8} and Table \ref{tab:table1App}). We show the trends of several power laws, $y\propto (h/h_{\rm{peak}})^{\alpha}$, for the normalized total flux, polarized flux, and integrated polarization fraction in Figure \ref{fig:fig8}. We find a negative slope with a general trend of $\alpha_{\rm{I}} \sim [-4,-3]$ and $\alpha_{\rm{PI}} \sim [-3,-2]$ for the total and polarized flux, respectively. The expected profiles for optically thin dust in total flux have slopes from $-2$ to $-4$ \citep{Leroy2015}. These trends suggest that the PI is flatter than I as a function of the distance across the galactic outflow.

 We compute the maximum vertical height of the polarized flux per galaxy and wavelength (Table \ref{tab:table1App}). The maximum extension of the outflows is found in the $89$--$214$ \um~wavelength range for all galaxies.  We estimate the maximum vertical height, $h_{\rm{max}}$, to be $1.9$\,kpc, $0.9$\,kpc, and $4.0$\,kpc for M\,82, NGC\,253, and NGC\,2146, respectively.  The $h_{\rm{max}} \sim 5$\,kpc at $214$\,\um~in NGC\,2146 corresponds to the polarization measurement in the west side of the galaxy, which is not spatially colocated with the outflow (Figures \ref{fig:fig1} and \ref{fig:fig2}). This polarization measurement may be part of the dust in the tidal tail due to the recent interaction that triggered star-formation in NGC\,2146 \citep{Martini2003,  Tarchi2004}.
 
 We find that the polarization fraction increases with the vertical height of the outflow (Figure \ref{fig:fig7} and \ref{fig:fig8}). Given the large differences in angular resolution between the JCMT and SOFIA observations and the ALMA observations, we separate the analysis of the polarization with angular resolutions $>5\arcsec$ (i.e., SOFIA and JCMT). We discuss the effect of the angular resolution in Section \ref{subsec:DIS_OBS}. We estimate the mean of the minimum polarization within the $50$--$850$ \um~wavelength range to be $\langle P^{\rm{int}}_{\rm{min}} \rangle = 0.8\pm0.5$\% with a range of $P^{\rm{int}}_{\rm{min}} = [0.2,1.8]$\%. The mean of the maximum polarization within the $50$--$850$\,\um~wavelength range is estimated to be $\langle P^{\rm{int}}_{\rm{max}} \rangle = 3.0\pm1.5$\% with a range of $P^{\rm{int}}_{\rm{max}} = [0.6,5.3]$\%. To study the variation of the fluxes alongside the vertical height, we normalize the polarization fraction to the peak at the maximum polarization of each band, $h/h_{\rm{max}}^{P^{\rm{int}}}$ (Figure \ref{fig:fig8}). We find a positive slope with a general trend of $\alpha_{\rm{P^{\rm{int}}}} \sim [1.5,2.5]$ for the integrated polarization. The polarization of the galactic outflows is measured up to vertical heights of $\sim4$\,kpc from the midplane of the galaxy at $89$ and $154$\,\um.

\subsection{Polarization spectra across the outflow}\label{subsec:PspecOut}

\begin{figure*}[ht!]
\includegraphics[width=\textwidth]{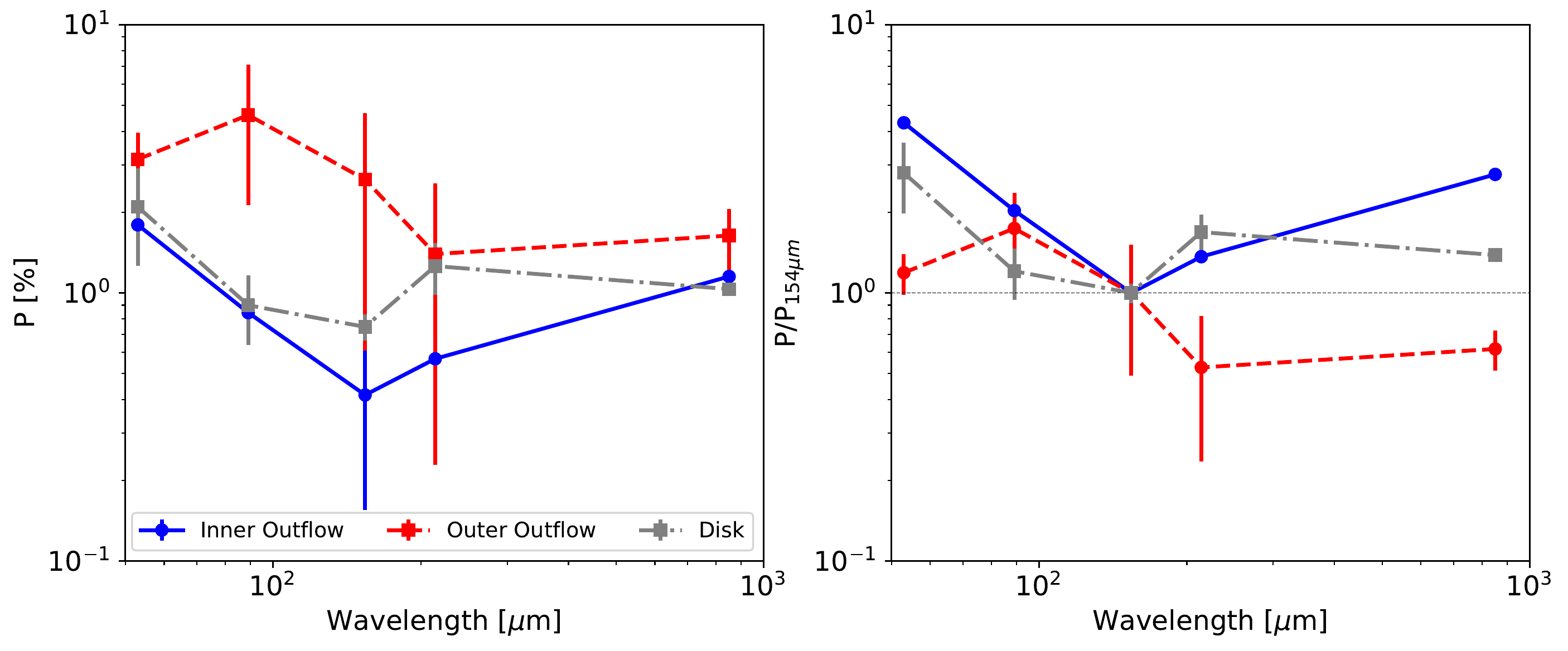}
\caption{Polarization spectra of the disk and outflow in starburst galaxies. The polarization spectrum (left) of the disk (grey), inner outflow (blue), and outer outflow (red) with the values in Table \ref{tab:table2} are shown. The normalized polarization spectrum (right) at a wavelength of $154$\,\um~is shown.
 \label{fig:fig9}}
\end{figure*}

We estimated the integrated polarization spectra of the disk and outflow of starburst galaxies (Section \ref{subsec:RES_P}).  However, we found that the polarization varies across the vertical height of the outflow (Section \ref{subsec:RES_r}). Based on the results from Section \ref{subsec:RES_r}, we now compute the polarization spectrum of the inner outflow as the integrated polarization within the maximum vertical height of the polarized flux (Table \ref{tab:table1App}). The polarization spectrum of the outer outflow is estimated as the integrated polarization after the maximum vertical height of the polarized flux. Figure \ref{fig:fig9} shows the polarization spectra of the disk, inner outflow, and outer outflow with their values shown in Table \ref{tab:table2}. Note the change between polarization spectra in Figures \ref{fig:fig6} and \ref{fig:fig9}, which shows the differences in the polarization spectra between considering the full galaxy and specific regions. This new set of polarization spectra differs from those presented in \citet[][SALSA IV]{SALSAIV} as now each physical component is disentangled and characterized independently. 

\begin{deluxetable}{lccccccccccccl}
\centering
\tablecaption{Polarimetric measurements of the inner and outer outflow, and disk of starburst galaxies. From left to right: 
a) Wavelength of the observations,
b) mean polarization fraction of the inner outflow,
c) mean polarization fraction of the outer outflow,
d) mean polarization fraction of the disk,
The $1\sigma$ uncertainty of b), c), and d) represent the dispersion of the polarization measurements at a given band, not the individual uncertainty of the polarization measurement. 
}
\label{tab:table2}
\tablewidth{0pt}
\tablehead{\colhead{Band}   &
 \colhead{P$_{\rm{InnerOutlflow}}$}  & 
 \colhead{P$_{\rm{OuterOutflow}}$} & 
  \colhead{P$_{\rm{Disk}}$} \\
\colhead{(\um)} & 
\colhead{(\%)} & 
\colhead{(\%)} & 
\colhead{(\%)}  \\
\colhead{(a)} & \colhead{(b)} & \colhead{(c)} & \colhead{(d)} } 
\startdata
53	&	$1.8\pm0.1$ 	&  $3.1\pm0.2$ & $2.1\pm0.8$ \\
89	&	$0.8\pm0.2$ 	&  $4.6\pm0.6$ & $0.9\pm0.3$ \\
154	&	$0.4\pm0.3$ 	&  $2.6\pm0.5$ & $0.7\pm0.2$ \\
214	&	$<0.6$ 		&  $1.4\pm0.3$ & $1.3\pm0.3$ \\
850	&	$<1.2$ 		&  $1.6\pm0.4$ & $1.0\pm0.2$ \\
\enddata
\end{deluxetable}

We characterize the $50$--$850$ \um~polarization spectra of the disk, outer outflow, and inner outflow separately. As mentioned in Sections \ref{subsec:RES_I} and \ref{subsec:RES_P}, the $890$\,\um~polarization measurement is not taken into account in this analysis because these observations trace different physical scales. The galactic disk has a fairly constant polarization fraction of $1.2\pm0.5$\% in the $53$--$850$\,\um~wavelength range. The high polarization fraction at $53$\,\um~is dominated by the outskirts regions of the disk of NGC~2146 (Figures \ref{fig:fig1} and \ref{fig:fig2}), and it may be interpreted as an upper limit. For the inner outflow, the polarization fraction falls from $1.8\pm0.1$\% at $53$\,\um~to a minimum of $0.4$\% in the $89$--$214$\,\um~wavelength range and then increases up to $\sim1.2$\% at $850$\,\um. The outer outflow has a peak in the polarized spectra of $4.6\pm0.6$\% within the $53-154$\,\um~wavelength range. Then, the polarization decreases down to $1.4$\% at $214$\,\um, and it seems to remain constant up to $850$\,\um.


\section{Discussion}\label{sec:DIS}

\subsection{Polarization spectrum of starburst galaxies}\label{subsec:DIS_P}

\begin{figure*}[ht!]
\includegraphics[width=\textwidth]{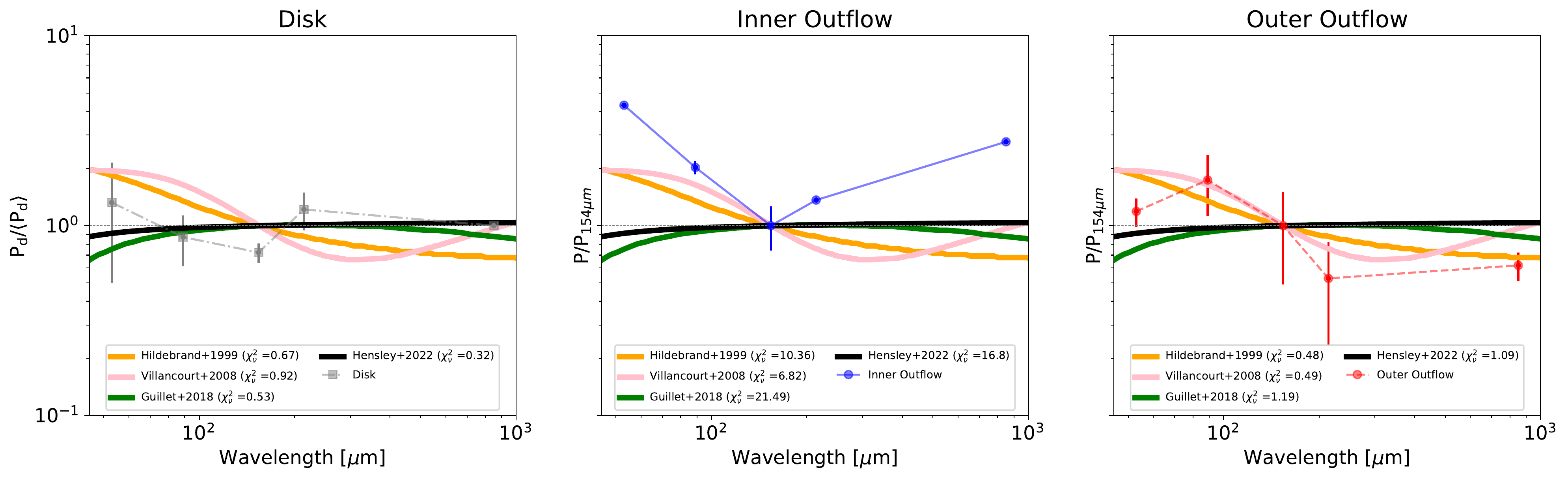}
\caption{Comparison of dust models and observed polarization spectra. The polarization spectra of the disk (left), inner outflow (middle), and outer outflow (right) as shown in Figure \ref{fig:fig9}. We show the polarization models for the diffuse ISM by \citet[fig. 13, G=$10^{3}$]{Guillet2018} (green) and \citet[fig.18, $U=10^{3}$]{Hensley2022} (black), and for a heterogeneous cloud with two temperature dust components by \citet[fig. 7]{H1999} (orange) and \citet[fig. 3]{Vaillancourt2008} (grey). The reduced $\chi_{\nu}^{2}$ for each model is shown.
 \label{fig:fig10}}
\end{figure*}

The polarization spectra of magnetically aligned dust grains have only been modeled thus far to understand the diffuse ISM, molecular clouds, and star-forming regions in the Galaxy. We summarize these models and study how they may be applied to the polarization spectra of starburst galaxies.

For the diffuse ISM (A$_{\rm{v}} < 2.5$ mag.), \citet{DF2009} models, based on \citet{DL2007}, assumed a mixture of spheroidal silicate and graphite grains to compute the polarized SED in the $2$--$3000$ \um~wavelength range. For all their dust composition configurations, these models \citep[fig. 8 by][]{DF2009} have polarization spectra that increase with increasing wavelength in the $50$--$850$\,\um~wavelength range. Recent studies using \textit{Planck} observations have revisited these models. For example, \citet{Guillet2018} modeled a combination of polycyclic aromatic hydrocarbons (PAHs), astrosilicates, and amorphous carbon grains, where only a fraction of amorphous carbon grains are aligned with the B-field, and astrosilicates are always aligned. Most of their models \citep[fig. 13 by][]{Guillet2018} show a polarization spectrum increasing with increasing wavelength in the $50-850$ \um~wavelength range. The polarization spectrum is flatter in the $50$--$850$\,\um~wavelength range with a peak at $\sim100$\,\um~only when the radiation field intensity ($G_{0}$) of the interstellar radiation field (ISRF) heating the dust grains is increased to values of  $G_{0}= 10^{3}$. $G_{0} = 1$ is defined as the intensity of the radiation field integrated between $6$ and $13.6$\,eV for the standard ISRF. Furthermore, \citet{DH2021} and \citet{Hensley2022} have put forward a new model of interstellar dust, the ``astrodust'' model, based on the phenomenology of dust in the diffuse Galactic ISM \citep{HD2021}. These models predict that the polarization fraction increases with increasing wavelength \citep[fig. 18,][]{Hensley2022} in the $50$--$850$ \um~wavelength range with an inflection point moving to shorter wavelengths as the radiation field intensity, $U$, increases. Note the different nomenclature referring to the radiation heating the dust, where $U\simeq1.6G_{0}$ by \citet{Hensley2022}. The dust models with high $G_{0}$ and $U$ are particularly interesting in our work because it represents the ISM under a strong radiation field from star-forming regions.

For molecular clouds and star-forming regions in the Galaxy, the polarization fraction has been observed to fall from $60$\,\um~to $350$\,\um, then rise from $350$\,\um~to $1300$\,\um~\citep[e.g.,][]{H1999,Vaillancourt2002,Vaillancourt2008,Gandilo2016,Ashton2018,Shariff2019,Michail2021}. \citet{H1999} found that dense cloud cores (e.g., Orion BN/KL) have a rising $50$--$1000$\,\um~polarization spectrum. The envelopes of the clouds OMC-1 and M\,17 have a falling $50$--$400$\,\um~polarization spectrum. The latter (i.e., falling polarization spectrum) is thought to be produced by a mix of dust grain composition with their efficiency changing as a function of dust temperature. These authors suggested that the regions with higher dust temperatures contain aligned dust grains. The former (i.e., rising polarization spectrum) agrees with an expected increase of polarization fraction as the optical depth decreases. This trend is also compatible with the models of the diffuse ISM. \citet{Vaillancourt2002} and \citet{Vaillancourt2008} computed the polarization spectrum using a two-component dust model assuming dust temperatures of $20$ and $50$\,K with the emissivity indexes of $\beta\,=\,2$ and 1, respectively. This model assumes that only the warm component is polarized. The resulting polarization spectrum falls within the $40$--$300$\,\um~wavelength and rises within the $300$--$2000$\,\um~wavelength range, with a minimum in the range of $100$--$350$\,\um.

Figure \ref{fig:fig10} shows the dust models of the polarization spectrum for strong radiation fields in the diffuse ISM; $G_{0}=10^{3}$ by \citet{Guillet2018} and $U=10^{3}$ by \citet{Hensley2022}. We also include the polarization spectrum of a heterogeneous cloud \citep[fig. 7 in][i.e, falling spectrum]{H1999} and the two-component dust temperature \citep[fig. 3 in][]{Vaillancourt2008}. We normalize the polarization spectrum of the disk to the mean of the polarization within the $50$--$850$\,\um~wavelength range in Figure \ref{fig:fig10} because the trend is not statistically significant, in contrast with the polarization spectrum of the inner and outer outflow. The polarization spectra of the inner and outer outflows are normalized to the wavelength of $154$\,\um. We estimate the reduced $\chi_{\nu}^{2}$ to compare the measured polarization spectra with the dust models, where $\chi_{\nu}^{2} = [(obs-model)^{2}/model^{2}]/(N-1)$ with $N$ the number of measurements. 

We find that the dust models of the diffuse ISM may be able to reproduce the polarization spectrum of the disk ($\chi_{\nu}^{2} = 0.32$ for \citet{Hensley2022} and $\chi_{\nu}^{2} = 0.53$ for \citet{Guillet2018}). This result suggests that the disk is characterized by having a single polarized dust component of optically thin dust that resembles similar dust properties of the diffuse ISM in the Galaxy. This model is also compatible with the estimated single dust temperature blackbody function from our fits to the total and polarized SEDs of the disk (Fig. \ref{fig:fig4}). However, the computed dust temperatures ($T_{\rm{d}} =[21,32]$\,K) using the SEDs and the estimated dust temperature ($T_{\rm{d}} \sim 70$\,K) from the models with $U=10^{3}$ seem to be incompatible. This difference may arise from a distribution of radiation fields in the measured polarization spectrum of the disk. The thermal emission from the disk is closely located at the base of the outflow (Figs. \ref{fig:fig1} and \ref{fig:fig2}). We expect that a certain fraction of the grains are heated with $U=10^{3}$, while another is heated with $U\sim1$. Note that most of the measured polarization at $<154$\,\um~is arising from the proximity of the base of the outflow, while the polarization from the disk is more prominent at $\ge154$\,\um. This effect is clearly evident in the polarized flux maps of NGC\,2146 (Fig. \ref{fig:fig2}): the polarized flux from the galactic outflow dominates at $53$--$89$\,\um, and it is completely unpolarized at $214$\,\um, at which the disk's polarized flux dominates. In addition, Figure \ref{fig:fig8} shows that the $53$--$89$\,\um~polarization spectrum of the disk has a similar steep falling trend to the polarization spectrum from the inner outflow. This trend may be produced by the polarization spectrum of NGC\,2146 (Fig. \ref{fig:fig6}). Based on these results, the $53$--$154$\,\um~polarization of the disk may be affected by the high polarization fraction from the strong radiation fields in the outflow producing a false-negative flat spectrum. With higher angular resolution observations, the polarization spectrum should show an increase in polarization or a flat polarization spectrum with $U<<10^{3}$ as expected for the diffuse ISM.

The heterogeneous cloud model and the two temperature dust models are the most favorable for explaining the polarization spectrum in the outer outflow ($\chi_{\nu}^{2} = 0.48-0.49$ for the \citet{H1999} and \citet{Vaillancourt2008}, respectively). These models are in agreement with our best-fits models of the total and polarized SEDs. We showed that the aligned dust grains producing the polarized SED have higher dust temperatures ($\sim 45$\,K) than the population of dust grains producing the total flux SED ($[31-41]$\,K). In addition, the aligned dust grains also have higher dust emissivities ($\beta \sim 2.3$) than the population of dust grains producing the total flux SED ($\beta \sim1.6$). These results suggest that the outer outflow has a population of warmer dust that is polarized while the colder dust is not polarized. 

The polarization spectrum of the inner outflow cannot be reproduced by any of the dust models of the diffuse ISM or dense clouds. The steep decrease of the polarization spectrum in the $50-154$\,\um~indicates large dust temperature gradients.  The increase of the polarization spectrum at $\lambda>154$\,\um~may indicate the transition from hot dust embedded in the starburst region to optically thin dust from a colder component along the LOS as wavelength increases. A multi-component dust model with larger dust temperature gradients may be required to explain the polarization spectrum of the inner outflow.
 
We conclude that, in the disk, the fraction of the dust population producing the polarized SED has similar physical conditions to the dust population contributing to the total flux SED. However, the measured polarization may be affected by a distribution of radiation fields producing a bias toward large polarization fractions at $<154$\,\um. In contrast, in the outflow, the polarized dust is embedded in a hotter environment than the dust producing the total flux SED. The shapes of the polarization spectra are due to variations in the dust temperature and dust grain emissivity. There may also be variations in dust composition and/or grain alignment efficiency that require further modeling.

\subsection{Polarized outflow and star formation rate}\label{subsec:DIS_SFRvsh}

\begin{figure}[ht!]
\includegraphics[width=\columnwidth]{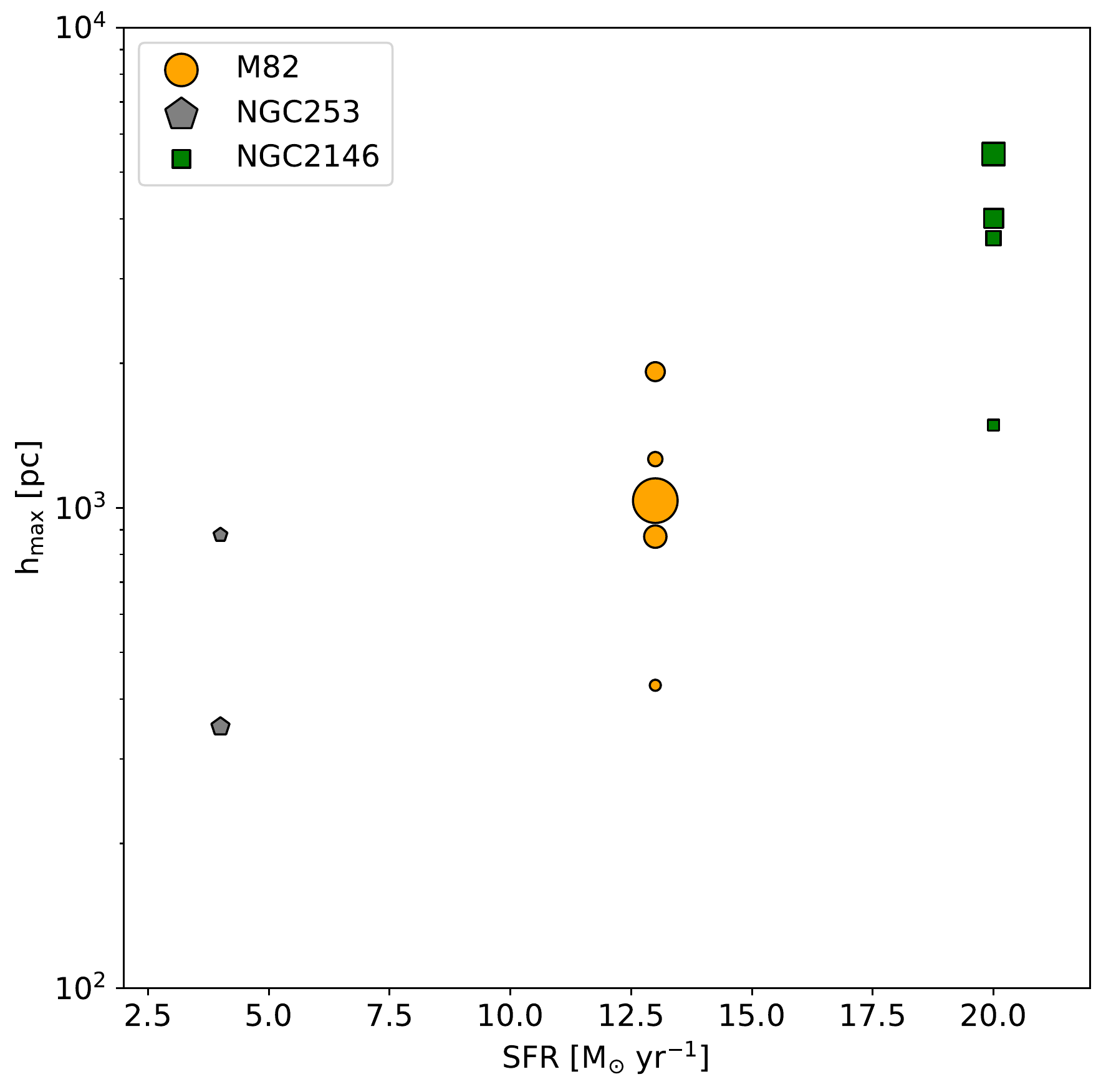}
\caption{Extension of the polarized intensity with the star formation rate. The maximum vertical height of the polarized flux, $h_{\rm{max}}$, as a function of the SFR (Table \ref{tab:table1}) per wavelength and galaxy. The size of the markers increases with wavelength. The maximum vertical heights from Table \ref{tab:table1App} are shown.
 \label{fig:fig11}}
\end{figure}

We analyze the extension of the galactic outflows in polarized intensity with the SFR of the starburst galaxies. Figure \ref{fig:fig11} shows that the maximum vertical height of the polarized intensity extends from $\sim0.3$\,kpc to $\sim4$\,kpc within the SFR $=[3,20]$\,M$_{\odot}$\,yr$^{-1}$ range, respectively. Using the polarized intensity as a proxy of the presence of the B-fields in the cold phase of the ISM, our observations show that the extension of the B-field permeating the CGM increases as a function of the SFR.

The B-field strength has been found to be correlated with the gas surface density and SFR. This correlation emerges from flux-freezing, and it can analytically be determined using the FIR-radio relation \citep{Schleicher2016}. Specifically, turbulence amplifies the B-field through fluctuation dynamo action \citep{Schober2012}, which converts turbulent kinetic energy into magnetic energy. As the SFR injects turbulent energy into the medium, this conversion couples the B-field strength with the SFR. The total B-field strength of nearby spiral galaxies has been found to be correlated with the star formation rate, $B_{\rm{tot}} \propto SFR^{0.34\pm0.04}$ \citep{VanEck2015,Beck2019}. These results confirm the theoretical prediction of $B_{\rm{tot}}\propto SFR^{1/3}$ using models of a galaxy dominated by supernova-driven turbulence, where the energy of the turbulent B-field is a fixed fraction of the turbulent energy \citep{Schober2013}. 

We estimate the expected turbulent B-field strength of the starburst galaxies in our sample using the $B\,=\,B_{\rm{0}}\,SFR^{0.34\pm0.04}$ relation. The turbulent magnetic strength was estimated to be $B_{\rm{M82}}\,=\,305\pm5\,\mu$G. Using M82's result and $SFR_{M82}\,=\,13$\,M$_{\odot}$\,yr$^{-1}$ (Table \ref{tab:table1}), the scaling factor is $B_{\rm{0}}\,=\,128\,\mu$G\,M$_{\odot}^{-0.34}$\,yr$^{-0.34}$. We compute the expected turbulent B-field strength of $B_{\rm{NGC253}} \sim 186~\mu$G and  $B_{\rm{NGC2146}}\,\sim\,354\,\mu$G for NGC\,253 and NGC\,2146, respectively. These expected B-field strengths are in good agreement with the revised equipartition between B-fields and cosmic rays taking into account energy losses due to strong B-fields in a dense medium put forward by \citet{LB2013}. These authors estimated equipartition B-field strengths of $240~\mu$G, $230~\mu$G, and $190~\mu$G for M\,82, NGC\,253, and NGC\,2146, respectively. Differences between methods may be due to the intrinsic nature of the tracer---FIR observations trace dense and cold ISM, while radio observations trace warm and diffuse ISM. Another difference may be the assumption of equipartition between cosmic rays and magnetic energy at radio wavelengths, and between turbulent kinetic energy and turbulent magnetic energy at FIR wavelengths. At FIR wavelengths, the B-fields have been measured to be more chaotic than those traced at radio wavelengths \citep{Surgent2023,SALSAV}. MHD turbulence simulations in a cube have also shown more tangled B-fields in the cold (T\,$<10^{3}$\,K) phase than in the warm (T\,$\ge10^{3}$\,K) phase \citep{SF2022}. These works may indicate that different levels of turbulent B-field strengths may be present in the multi-phase ISM. In addition, there are severe energy losses due to the cosmic rays propagating in a very dense medium with strong B-fields \citep{LB2013}. These energy losses may decrease the measured B-field strengths from those computed at FIR wavelengths.

We compute the energy budget to quantify the role of the B-fields in the outflow of starburst galaxies. Let the turbulent kinetic energy, $U_{\rm{K}}$, and turbulent magnetic energy, $U_{B}$, be 

\begin{eqnarray}
    U_{\rm{K}} &=& \frac{1}{2} \rho \sigma_{\rm{v}}^{2} \\
    U_{\rm{B}} &=& \frac{B^{2}}{8\pi}
\end{eqnarray}
\noindent 
where $\rho$ is the volume density, $\sigma_{\rm{v}}$ is the dispersion velocity, and $B$ is the B-field strength. The volume density was estimated as $\rho=\Sigma_{\rm{g}}/h_{\rm{c}}$, where $\Sigma_{\rm{g}}$ is the surface gas density and $h_{\rm{c}}$ is the depth of the starburst region. 

\begin{deluxetable*}{lcccccccl}
\centering
\tablecaption{Expected energy budget across the galactic outflow. From left to right: 
a) Object,
b) B-field strength at the core,
c) size of the core,
d) surface gas density,
e) velocity dispersion of the molecular gas,
f) turbulent magnetic energy,
g) turbulent kinetic energy,
h) $\beta^{\prime}$-plasma parameter, $\beta^{\prime} = U_{\rm{K}}/U_{\rm{B}}$,
i) references.
}
\label{tab:table3}
\tablewidth{0pt}
\tablehead{\colhead{Object}   &
 \colhead{B}  & 
 \colhead{h$_{\rm{c}}$} & 
  \colhead{$\Sigma_{\rm{g}}^{\dagger}$} & 
  \colhead{$\sigma_{\rm{v}}^{(1)}$}  & 
  \colhead{U$_{\rm{B}}$} & 
  \colhead{U$_{\rm{K}}$} & 
  \colhead{$\beta^{\prime}$} &
  \colhead{References} \\
\colhead{} & 
\colhead{($\mu$G)} & 
\colhead{(pc)} & 
\colhead{(g cm$^{-2}$)} & 
\colhead{(km s$^{-2}$)} & 
\colhead{($\times10^{9}$ g cm$^{-1}$ s$^{-2}$)} & 
\colhead{($\times10^{9}$ g cm$^{-1}$ s$^{-2}$)} & \\
\colhead{(a)} & \colhead{(b)} & \colhead{(c)} & \colhead{(d)} & \colhead{(e)} & \colhead{(f)} & \colhead{(g)} & \colhead{(h)} & \colhead{(i)}} 
\startdata
M82 			&	305$^{\dagger\dagger}$	&	500		&	0.69	&	66.6	&	3.70	&	9.92	& 2.68	&	$^{(1)}$\citet{Leroy2015}	\\
NGC~253		&	186	&	50		&	0.47	&	50	&	1.35	&	38.1	&	28.2	&	$^{(1)}$\citet{Krieger2019}	\\
NGC~2146	&	354	&	2000		&	0.12	&	250	&	4.93	&	7.08	&	1.43	&	$^{(1)}$\citet{Kreckel2014} \\
\enddata
\tablecomments{$^{\dagger}$Surface gas densities were taken from \citet{LB2013}. $^{\dagger\dagger}$Measured B-field strength in the central $\sim1$ kpc by \citet{LR2021}.}
\end{deluxetable*}

We took $\Sigma_{\rm{g}}$ from \citet{LB2013}, and we assume that the depth of the starburst region of each galaxy is isotropically distributed in a disk with a depth equal to the length of the polarized region along the major axis of the galaxy from our observations (Figure \ref{fig:fig2}). We took the velocity dispersion of the molecular gas, CO, as a tracer of the cold and molecular outflow from the references shown in Table \ref{tab:table3}. Table \ref{tab:table3} shows these values and the estimated total turbulent kinetic and magnetic energy in the starburst region. For all starburst galaxies, the central $\sim 1$ kpc galactic outflow is in close equipartition, $\beta^{\prime} = U_{\rm{K}}/U_{\rm{B}} = [1.4, 28]$, between the turbulent kinetic and magnetic energies. We find that the turbulent magnetic energy becomes more dominant as the SFR increases.

\citet{LR2021} characterized the energies across the galactic outflow of M\,82 at $53$\,\um~using SOFIA/HAWC+. The galactic outflow of M\,82 was estimated to be close to equipartition ($\beta^{\prime}=0.56\pm0.23$) between the turbulent kinetic and magnetic energies within the central $\sim1$ kpc using direct measurements, and up to $6.6$\,kpc using a potential field extrapolation. Note that here we use the integrated surface density of the galaxy and assume that it is concentrated within the outflow region, which overestimates the volume density and therefore the turbulent kinetic energy. This is evident in the estimated $\beta^{\prime} = 2.68$ in Table \ref{tab:table3} and the estimated $\beta^{\prime}=0.56\pm0.23$ using resolved observations and the more sophisticated approach to estimate the B-field strengths of M\,82. In addition, \citet{LR2021} showed that the energy equipartition produces `open' B-field lines into the CGM. The magnetic energy was measured to dominate in the interclump medium up to a distance of $6.6$ kpc-scales. Here, we show the direct observations of the B-field lines in M\,82 up to $\sim2$\,kpc and $\sim4$\,kpc in NGC~2146, which extend the analysis to SFR in the range of $[3-20]$\,M$_{\odot}$\,yr$^{-1}$. 

Our analysis suggests that a) the B-field becomes stronger as the SFR increases, b) the magnetic surface density profile seems to be flatter than the total density profile of the outflow (Sec. \ref{subsec:RES_r}), and c) the B-fields are dragged from the disk to the galactic outflows and may be `open' into the CGM. 

There are several caveats to the method presented in this section. The B-field strength at large vertical heights is currently assumed to be unscreened. It is unknown how the geometry and strength of the B-fields propagate across the heterogenous distribution of gas and dust in the outflow. The observations of NGC~2146 are the only ones showing that the B-field is still parallel to the galactic outflow at scales of $\sim4$ kpc from the disk. In addition, the total B-field strength is typically used in the relation between the SFR and the B-field strength, while here we used the turbulent B-field strength. The total B-field strength is  $1.4-4.2$ times larger than the ordered B-field \citep{Beck2019}, which implies that $\beta^{\prime}$ would be $<1$ (i.e., magnetically dominated outflows) at the core of starburst galaxies than those estimated here. Finally, the SFR and total B-field strength relation has been derived using spiral galaxies, which implies that this relationship has to be revisited due to energy losses in the dense ISM and strong B-fields found in starburst galaxies. 

Nevertheless, the results presented here show that strong astrophysical B-fields amplified by the starburst activity are pushed away into the CGM. These B-fields have a flatter radial profile across the outflow than the total intensity (Fig. \ref{fig:fig9}), which indicates that B-fields may become more dynamically important in the CGM than the kinetic energy of the galactic outflow.

\subsection{Observational strategies}\label{subsec:DIS_OBS}

\begin{figure*}[ht!]
\includegraphics[width=\textwidth]{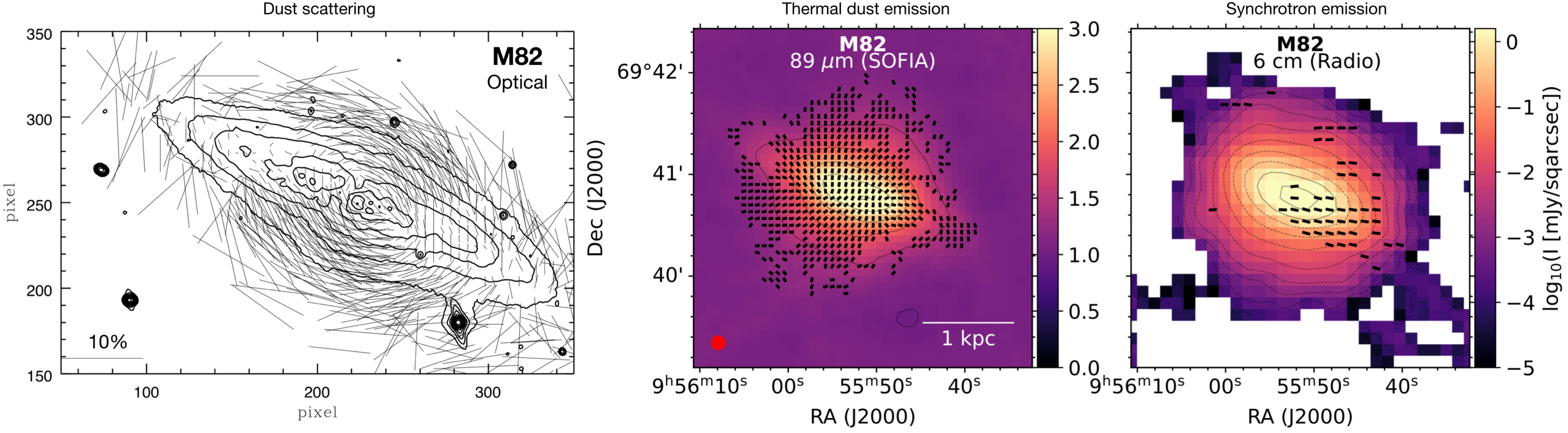}
\caption{In search of the B-field signature along the galactic outflows of starburst galaxies: M\,82 as the case study. Optical polarimetric observations are dominated by dust scattering (left) \citep{Neininger1990,Fendt1998}. The lines show the E-vectors with the 10\% polarization legend shown in the bottom-left. FIR polarimetry observations reliably trace the B-fields in the galactic outflows (middle) \citep{Jones2019,LR2021,SALSAV}. Radio polarimetric observations have short lifetime cosmic rays along the galactic outflow, and only a magnetized bar is measured (right) \citep[e.g.,][]{Thompson2006,Adebahr2017}. For both FIR and radio, the lines show the B-field orientation with constant length. 
 \label{fig:fig12}}
\end{figure*}

As mentioned in the Introduction (Section \ref{sec:INT}), Optical/NIR polarimetric observations are dominated by dust and/or electron scattering and do not provide information about B-fields in starburst galaxies. Radio polarimetric observations may have energy losses due to the high B-field strengths, the short lifetime of the CR \citep{Thompson2006}, and Faraday rotation effects \citep{Beck2019}. The B-fields in the galactic outflows may be very challenging to observe with radio polarimetric observations. The $50$--$850$\,\um~polarimetric observations presented here have shown to be excellent tracers of the B-fields in the outflows of starburst galaxies. Figure \ref{fig:fig12} shows the optical, FIR, and radio polarimetric observations of M\,82 as the showcase for the search of B-fields using multi-wavelength polarimetric observations. Our analysis has shown that the polarization properties of the galactic outflow can be disentangled from those in the galactic disk of starburst galaxies using $53$--$850$\,\um~polarimetric observations. Galactic outflows are better traced at wavelengths $<154$\,\um, while the disk is better traced at $\lambda>154$\,\um~at angular resolution of $5$--$18\arcsec$ ($84$--$1500$\,pc).

The extension of the polarized emission of the galactic outflow varies with wavelength (Section \ref{subsec:RES_r}), the sensitivity of the instruments, and the dust temperature gradient in the galactic outflow. For the latter, the galactic outflows have their polarized dust mainly located in hotter dust than the total intensity (Fig.\,\ref{fig:fig4}) and with larger extensions (Figures \ref{fig:fig2} and \ref{fig:fig7}) than the cold dust temperatures traced at longer wavelengths. For the former, SOFIA/HAWC+ has been shown to better recover the large-scale extended total and polarized emission of the starburst galaxies than those observations from JCMT/POL-2 and ALMA. \citet{Pattle2021} suggested that the galactic outflow breaks out at $\sim350$\,pc above the disk. However, the $53$--$214$\,\um~wavelength polarimetric observations show that the B-field extends up to $2$\,kpc above and below the disk of M\,82. The JCMT/POL-2 result may be due to a combination of shallow polarimetric observations, multi-temperature components at $850$\,\um, and loss of large-scale extended emission.  Although the JCMT/POL-2 observations suffer from loss of large-scale extended emission in the total flux (Figure \ref{fig:fig4}), the polarized flux is fully recovered at the base of the outflow. This result implies that the polarized flux of the cold dust is fully captured within the central $400$\,pc of M82. Figure\,\ref{fig:fig8} shows that the polarized extended emission at $850$\,\um~decreases faster than those at shorter wavelengths. In addition, the polarization from the disk becomes more important at larger wavelengths because of the larger contribution of the cold dust component in the host galaxy than in the outflow. These results, in combination with the low SNR from the JCMT/POL-2 observations, cause a non-detection of the polarized dust in the outflow at $850$\,\um. As mentioned above, the galactic outflows have their polarized dust mainly located at hotter temperatures, which produces a smaller galactic outflow in polarized flux at longer wavelengths. The smaller B-field outflow region observed at $850$\,\um~is therefore an effect of the dust temperature gradient traced by the single observed wavelength. The maximum extension of the polarized outflows should be characterized using $\sim50$--$200$\,\um~polarimetric observations.

\begin{figure}[ht!]
\includegraphics[width=\columnwidth]{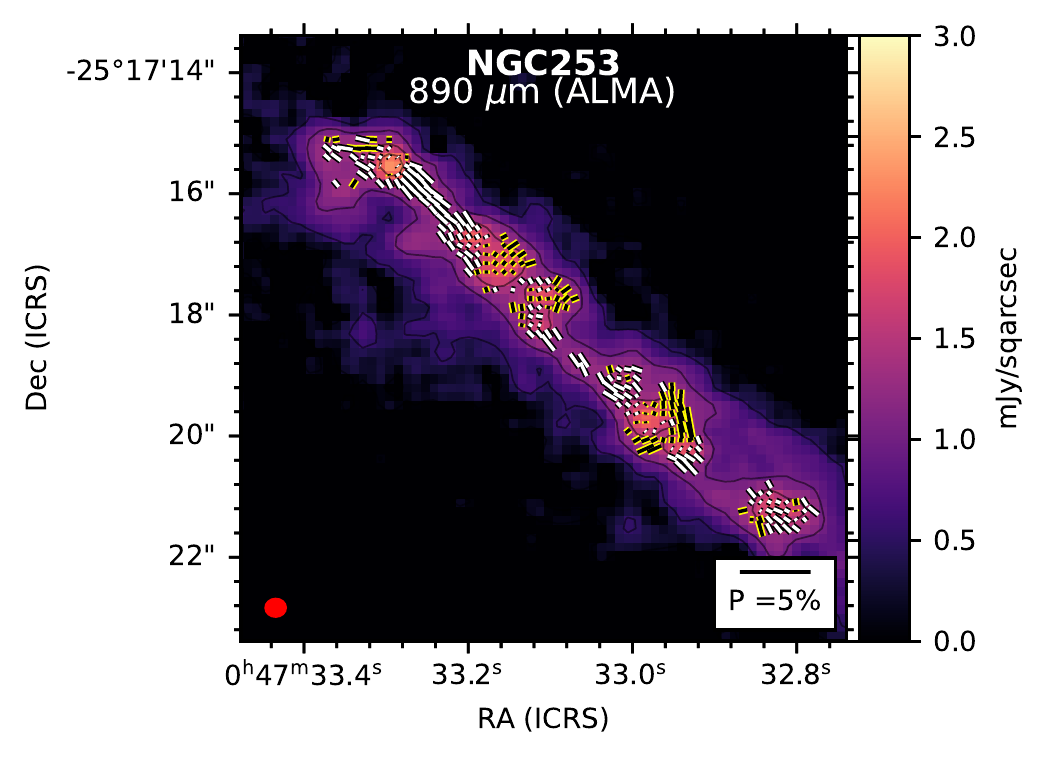}
\caption{The B-field orientation of the central $150$\,pc of NGC\,253. Zoom-in of the panel shown in Fig. \ref{fig:fig1}. 
 \label{fig:fig13}}
\end{figure}

The ALMA polarimetric observations at $890$\,\um~of NGC\,253 (Belfiori, D. et al. in preparation) revealed the fine structure of the B-fields at the highest spatial resolution ($5$\,pc) from our sample. The starburst region has a filamentary B-field structure parallel to the disk of the galaxy connecting the star-forming clusters, whereas the B-field in the star-forming clusters is perpendicular to the disk (Figure \ref{fig:fig13}). The B-field associated with the outflow is marginally observed at FIR in the central $300$\,pc. As shown in Figure \ref{fig:fig11}, the extension of the polarized flux depends on the SFR of the starbursts. Although there may be an extended B-field in the ALMA polarimetric observations, this large-scale B-field component is lost due to the interferometric mode. Further ALMA polarimetric observations at a lower angular resolution ($\sim1$\arcsec) are required to recover the B-field at the $50-300$ pc scales. These high-spatial resolution observations are required to resolve the turbulent coherence length of the B-field. These observations can be used to compute the B-field strength using a similar approach to that applied to M82 \citep{LR2021}, after the thermal and non-thermal component from the star-forming regions is disentangled at sub-mm wavelengths.


\section{Conclusions}\label{sec:CON}

We have presented a multi-wavelength, $53$--$890$\,\um, imaging polarimetric analysis of the nearby, $3.5$--$17.20$\,Mpc, starburst galaxies M\,82, NGC\,253, and NGC\,2146. These galaxies were observed using SOFIA/HAWC+, JCMT/POL-2, and ALMA at angular resolutions of $4.85$--$18.2\arcsec$, $14\arcsec$, and $\sim0.3\arcsec$, respectively. The measured polarization arises from thermal polarized emission by magnetically aligned dust grains, which provides the B-field orientation on the plane of the sky. For all galaxies, we resolved, disentangled, and characterized the B-field orientation in the disk and outflow. 

The B-field in the disk and outflow are disentangled using a geometric analysis (Section \ref{subsec:MET_PA}, Figure \ref{fig:fig1}). We computed the total and polarized SEDs and characterized them using a modified blackbody function (Section \ref{subsec:MET_SEDFitting}, Figure \ref{fig:fig4}). We found that the total flux and polarized flux SEDs of the disk are characterized by having similar low dust temperatures, $T_{\rm{d,disk}} = [24,31]$\,K, and dust emissivities of $\beta_{\rm{disk}} \sim1$. The outflow has different dust populations in the total flux and polarized flux SEDs. The total flux SED is characterized by having $T_{\rm{d,outflow}}^{I} = [31,41]$\,K and $\beta_{\rm{d,outflow}}^{I} \sim1.5$, while the polarized SED has higher dust temperatures $T_{\rm{d,outflow}}^{PI} \sim 45$\,K and $\beta_{\rm{d,outflow}}^{PI} \sim2.3$. This result implies that the polarized SED in the outflow arises from a dust grain population with higher dust temperature and dust emissivities than that from the total flux SED. In contrast, the same dust grain population in the disk produces the total and polarized SEDs. If galactic outflows are not resolved, these results show that the polarized SEDs of starburst can be used to better distinguish between disk-dominated and outflow-dominated galaxies than the total flux SEDs.

We computed the polarization spectra of the disk and outflow. We found that the polarization spectrum of the disk is mainly flat with a mean polarization fraction of $\langle P_{\rm{disk}} \rangle = 1.2\pm0.5$\% in the $53-850$\,\um~wavelength range. The polarization spectrum of the inner outflow (i.e., polarized flux within the vertical height of the galaxy disk) falls from $1.8\pm0.1$\%. to $0.4\pm0.3$\% in the $53$--$154$\,\um~and then rises up to $1.2$\% at $850$\,\um. The polarization spectrum in the outer outflow (i.e., polarized flux outside the vertical height of the disk) has a minimum in the $89$--$154$\,\um~with a peak of polarization of $4.6\pm0.6$\% at $89$\,\um. 

Although there are no dust models to work directly with the polarization arising from starburst galaxies, we compare our result with the dust models of the diffuse ISM and those from star-forming regions in the Galaxy. We estimated that the most likely dust polarization configuration for the disk is that from the diffuse ISM arising from a single dust temperature component. The recent `astrodust' models \citep{HD2021,Hensley2022} with strong radiation fields, $U=10^{3}$, seem to best reproduce the polarization spectrum of the disk. The outer outflow polarization spectrum is best reproduced by the models of heterogenous clouds \citep{H1999} and two-temperature dust components \citep{Vaillancourt2008}. These results suggest that the polarization spectrum should be analyzed using several dust components to fit simultaneously the total and polarized SEDs. 

We analyzed the polarization properties of the galactic outflows as a function of the vertical height. We found that the polarized flux extends from $0.3$\,kpc up to $\sim4$\,kpc with a maximum extension in the $89$--$154$\,\um~wavelength range. The polarized flux surface density drops with a general slope of $\sim[-3,-2]$, while the total flux decreases with a slope of $\sim[-4,-3]$. We found that the polarized flux extension of the galactic outflow increases with the global SFR of the galaxy with their central $\sim1$\,kpc in close equipartition between the turbulent kinetic and magnetic energies. Our analysis shows that the B-fields from the galactic disk can be dragged to extensions of $\sim4$\,kpc in the CGM by starburst galaxies with global SFR of $3-20$\,M$_{\odot}$\,yr$^{-1}$.


\begin{acknowledgments}
E.L.R. thanks Brandon Hensley for the great insights about the dust models, and Annie Hughes, Rosita Paladino, and Davide Belfiori for the discussions about the ALMA polarimetric observations of NGC\,253. 
Based on observations made with the NASA/DLR Stratospheric Observatory for Infrared Astronomy (SOFIA) under the 05\_0071, 08\_0012, and 07\_0032 Programs. SOFIA is jointly operated by the Universities Space Research Association, Inc. (USRA), under NASA contract NNA17BF53C, and the Deutsches SOFIA Institut (DSI) under DLR contract 50 OK 0901 to the University of Stuttgart. 
This paper makes use of the following ALMA data: ADS/JAO.ALMA\#2018.1.01358.S. ALMA is a partnership of ESO (representing its member states), NSF (USA) and NINS (Japan), together with NRC (Canada), MOST and ASIAA (Taiwan), and KASI (Republic of Korea), in cooperation with the Republic of Chile. The Joint ALMA Observatory is operated by ESO, AUI/NRAO and NAOJ. The National Radio Astronomy Observatory is a facility of the National Science Foundation operated under cooperative agreement by Associated Universities, Inc.

\end{acknowledgments}


\appendix

\section{Stokes $Q$ and $U$ of individual galaxies}\label{App:StokesQU}

\begin{figure*}[ht!]
\includegraphics[width=\textwidth]{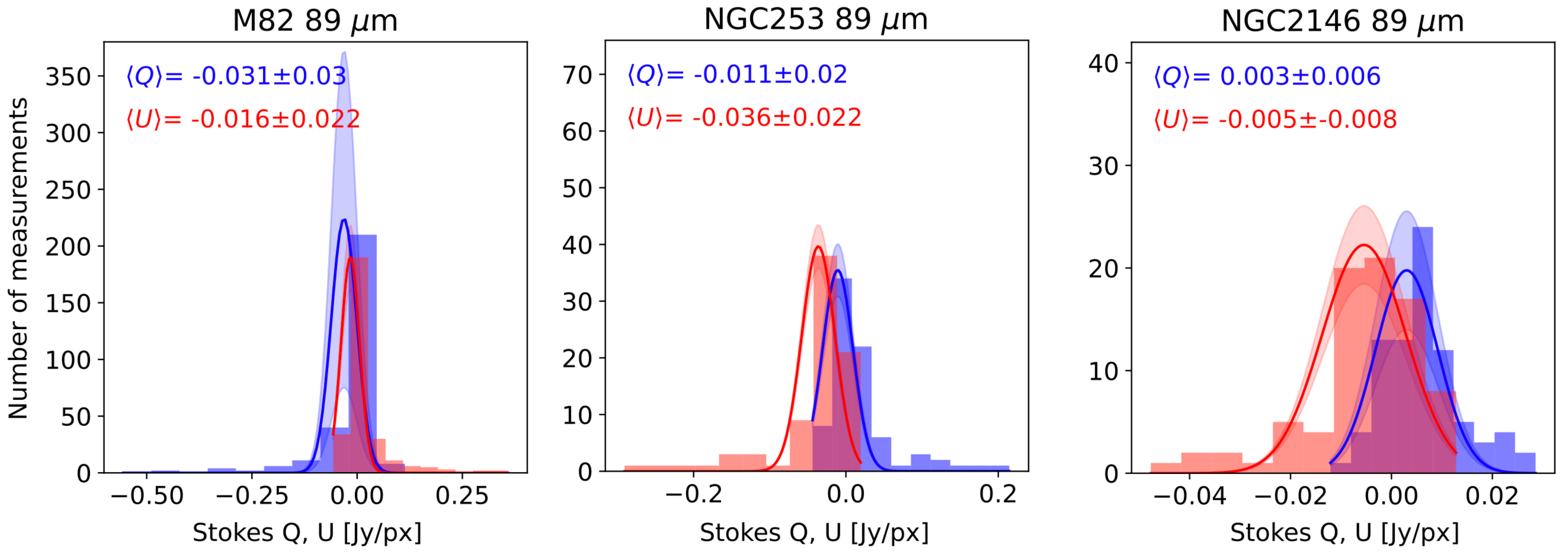}
\caption{Histograms of the Stokes $Q$ and $U$ of the starburst galaxies. The histograms of the Stokes $Q$ (blue) and $U$ (red) are fitted using a Gaussian profile. The best-fit mean (solid line) and $1\sigma$ (shadowed region) of the Gaussian profiles are shown in each plot. 
 \label{fig:fig1App}}
\end{figure*}

The histograms of the Stokes $Q$ and $U$ of the bands used to identify the outflow and disk of the starburst galaxies are shown in Figure \ref{fig:fig1App}. We use the polarization observations at $89$\,\um~of M\,82, NGC\,253, and  NGC\,2146. The histograms of the Stokes $Q$ and $U$ are fitted using a Gaussian profile with the mean and standard deviation as shown in each panel. The steps of this procedure are described in Section \ref{subsec:MET_PA}.

\section{Tabulated polarization measurements}\label{App:Pmes}

Table \ref{tab:table1App} shows the polarimetric measurements of the outflow and disk in starburst galaxies computed in the main text of this manuscript.

\begin{deluxetable*}{lcccccccccccccl}
\centering
\tablecaption{Polarimetric measurements of the outflow and disk in starburst galaxies. From left to right: 
a) Galaxy name,
b) wavelength of the observations,
c) mean polarization fraction of the outflow from the histograms,
d) mean polarization fraction of the disk from the histograms,
e) integrated polarization fraction of the full galaxy,
f) integrated polarization fraction of the outflow,
g) integrated polarization fraction of the disk,
h) vertical height at the peak of the polarized flux,
i) maximum vertical height,
j) maximum integrated polarization fraction,
k) minimum integrated polarization fraction.
The $1\sigma$ uncertainty of c) and d) represent the dispersion of the histograms, not the individual uncertainty of the polarization measurement. 
The $1\sigma$ uncertainty of c) and d) represent the uncertainty of the polarization measurement}
\label{tab:table1App}
\tablewidth{0pt}
\tablehead{\colhead{Galaxy}  & 	
\colhead{Band}   &
 \colhead{$\langle P^{\rm{hist}}_{\rm{outflow}}\rangle$}  & 
 \colhead{$\langle P^{\rm{hist}}_{\rm{disk}}\rangle$} & 
  \colhead{$P^{\rm{int}}_{\rm{galaxy}}$} &
  \colhead{$P^{\rm{int}}_{\rm{outflow}}$} &
   \colhead{$P^{\rm{int}}_{\rm{disk}}$} &
   \colhead{$h_{\rm{peak}}^{PI}$}  &
   \colhead{$h_{\rm{max}}$}  & 
  \colhead{$P^{\rm{int}}_{\rm{max}}$} &
  \colhead{$P^{\rm{int}}_{\rm{min}}$} \\ 
 		&
\colhead{(\um)} & 
\colhead{(\%)} & 
\colhead{(\%)} & 
\colhead{(\%)} &
\colhead{(\%)} &
\colhead{(\%)}  & 
\colhead{(pc)} & 
\colhead{(pc)} & 
\colhead{(\%)} & 
\colhead{(\%)} \\
\colhead{(a)} & \colhead{(b)} & \colhead{(c)} & \colhead{(d)} & \colhead{(e)} & \colhead{(f)} & \colhead{(g)} & \colhead{(h)} & \colhead{(i)} & \colhead{(j)} & \colhead{(k)} } 
\startdata
M82   		&	53	&	$3.2\pm1.1$	&	$2.2\pm1.4$	&  $1.6\pm0.4$	&  $1.9\pm0.4$	&  $1.3\pm0.4$	&	$143$	&	$427$	&	$3.2$	&	$1.8$	\\
			&	89	&	$2.6\pm1.7$	&	$1.9\pm2.1$	&  $0.9\pm0.4$	&  $1.1\pm0.4$	&  $0.9\pm0.4$	&	$223$	&	$1263$	&	$5.3$	&	$1.0$	\\
			&	154	&	$1.0\pm1.6$	&	$1.9\pm1.4$	&  $0.1\pm0.4$	&  $0.4\pm0.4$	&  $0.9\pm0.4$	&	$384$	&	$1921$	&	$3.5$	&	$0.3$	\\
			&	214	&	$0.7\pm1.0$	&	$2.0\pm1.7$	&  $0.2\pm0.4$	&  $0.5\pm0.4$	&  $1.5\pm0.4$	&	$523$	&	$871$	&	$0.6$	&	$0.4$	\\
			&	850	&	$1.1\pm1.3$	&	$1.9\pm2.0$	&  $0.4\pm0.5$	&  $0.8\pm0.5$	&  $1.0\pm0.5$	&	$444$	&	$1035$	&	$1.6$	&	$0.7$	\\
NGC~253   	&	89	&	$2.1\pm1.5$	&	$1.0\pm1.1$	&  $0.5\pm0.4$	&  $1.4\pm0.4$	&  $0.6\pm0.4$	&	$338$	&	$878$	&	$5.2$	&	$0.7$	\\
			&	154	&	$1.5\pm0.9$	&	$0.9\pm0.7$	&  $0.7\pm0.4$	&  $1.0\pm0.4$	&  $0.7\pm0.4$	&	$350$	&	$350$	&	$1.1$	&	$0.8$	\\
			&	890	&	$0.4\pm0.3$	&	$0.5\pm0.4$	&  $0.3\pm0.1$	&  $0.2\pm0.1$	&  $0.5\pm0.1$	&	$12$		&	$17$		&	$0.4$	&	$0.2$	\\
NGC~2146   	&	53	&	$2.0\pm1.6$	&	$2.8\pm0.5$	&  $1.7\pm0.4$	&  $1.7\pm0.4$	&  $2.9\pm0.4$	&	$637$	&	$1486$	&	$3.0$	&	$1.6$	\\
   			&	89	&	$1.0\pm1.4$	&	$1.5\pm1.7$	&  $0.5\pm0.4$	&  $0.6\pm0.4$	&  $1.2\pm0.4$	&	$996$	&	$3653$	&	$3.4$	&	$0.5$	\\
   			&	154	&	$0.6\pm2.2$	&	$1.0\pm1.5$	&  $0.3\pm0.4$	&  $0.3\pm0.4$	&  $0.7\pm0.4$	&	$573$	&	$4007$	&	$3.4$	&	$0.2$	\\
   			&	214	&	$1.5\pm1.5$	&	$1.2\pm1.2$	&  $0.9\pm0.4$	&  $1.3\pm0.4$	&  $1.0\pm0.4$	&	$3895$	&	$5453$	&	$2.3$	&	$0.7$	\\
			\enddata
\end{deluxetable*}

%

\vspace{5mm}
\facilities{SOFIA (HAWC+)}


\software{aplpy \citep{aplpy2012,aplpy2019},  
          astropy \citep{astropy:2013,astropy:2018,astropy:2022},
          pymc3 \citep{pymc}}

\bibliography{references}{}
\bibliographystyle{aasjournal}



\end{document}